
\let\includefigures=\iftrue

\input harvmac

\includefigures
\message{If you do not have epsf.tex (to include figures),}
\message{change the option at the top of the tex file.}
\input epsf
\def\figin{\epsfcheck\figin}\def\figins{\epsfcheck\figins}
\def\epsfcheck{\ifx\epsfbox\UnDeFiNeD
\message{(NO epsf.tex, FIGURES WILL BE IGNORED)}
\gdef\figin##1{\vskip2in}\gdef\figins##1{\hskip.5in}
\else\message{(FIGURES WILL BE INCLUDED)}%
\gdef\figin##1{##1}\gdef\figins##1{##1}\fi}
\def\DefWarn#1{}
\def\figinsert{\goodbreak\midinsert}
\def\ifig#1#2#3{\DefWarn#1\xdef#1{fig.~\the\figno}
\writedef{#1\leftbracket fig.\noexpand~\the\figno}%
\figinsert\figin{\centerline{#3}}\medskip\centerline{\vbox{
\baselineskip12pt\advance\hsize by -1truein
\noindent\footnotefont{\bf Fig.~\the\figno:} #2}}
\bigskip\endinsert\global\advance\figno by1}
\else
\def\ifig#1#2#3{\xdef#1{fig.~\the\figno}
\writedef{#1\leftbracket fig.\noexpand~\the\figno}%
\global\advance\figno by1}
\fi

\lref\ggl{H. Georgi and S.L. Glashow, Phys. Rev. Lett. {\bf 32} (1974) 438.}%
\lref\dg{S. Dimopoulos and H. Georgi, Nucl. Phys. {\bf B192}
(1981) 150.}%
\lref\RSAdS{J. Maldacena, unpublished; E. Witten, unpublished;  S.
Gubser, hep-th/9912001;
M. Duff and J. Liu, Phys. Rev. Lett. {\bf 85} (2000) 2052,
hep-th/0003237; N. Arkani-Hamed, M. Porrati and L. Randall,
hep-th/0012148; R. Rattazzi and A. Zaffaroni, hep-th/0012248. }

\lref\hs{M. Dine, W. Fischler, Nucl. Phys. {\bf B204} (1982) 346;
S. Dimopoulos and S. Raby, Nucl. Phys. {\bf B219} (1982) 479; J.
Polchinski and L. Susskind, Phys. Rev. {\bf D26} (1982) 3661.}
\lref\drw{S. Dimopoulos, S. Raby and F. Wilczek, Phys. Rev. {\bf
D24} (1981) 1681.} \lref\noscale{J. Ellis, C. Kounnas and D.
Nanopoulos, Nucl. Phys. {\bf B247} (1984) 373.} \lref\pomarol{A.
Pomarol, Phys. Lett. {\bf B486} (2000) 153; Phys. Rev. Lett. {\bf
85} (2000) 4004; T. Gherghetta and A. Pomarol, Nucl. Phys. {\bf
B586} (2000) 141, hep-ph/0012378.} \lref\suGUT{ E. Cremmer, S.
Ferrara, L. Girardello, and A. Van Proyen, Phys. Lett. {\bf B116}
(1982) 231; A. Chamseddine, R. Arnowitt, and P. Nath, Phys. Rev.
Lett. {\bf 49} (1982) 970; R. Barbieri, S. Ferrara, and C. Savoy,
Phys. Lett. {\bf B110} (1982) 343; L. J. Hall, J. Lykken, and S.
Weinberg, Phys. Rev. {\bf D27}(1983) 2359.}

\lref\rszero{L. Randall  and R. Sundrum, Nucl. Phys. {\bf B557}
(1999) 79; G.F. Giudice, M.A. Luty, H. Murayama and R. Rattazzi,
JHEP {\bf 9812} (1998) 027.}

\lref\nil{H.P. Nilles, Phys. Lett. {\bf B115} (1982) 193; Nucl.
Phys. {\bf B217} (1983) 366.} \lref\kaplu{V.S. Kaplunovsky and J.
Louis, Phys. Lett. {\bf B306} (1993) 269.}

\lref\gauco{M. Dine, R. Rohm, N. Seiberg, E. Witten, Phys. Lett.
{\bf B156} (1985) 55.}

\lref\juan{J. Maldacena, ATMP {\bf 2} (1998) 231, hep-th/9711200;
S. Gubser, I. Klebanov and A. Polyakov, Phys. Lett. {\bf B428} (1998)
105, hep-th/9802109; E. Witten, ATMP {\bf 2} (1998) 253, hep-th/9802150.}

\lref\RSI{L. Randall and R. Sundrum, Phys. Rev. Lett. {\bf 83}
(1999) 3370, hep-th/9905221.} \lref\AdSCFT{S. Gubser, I. Klebanov
and A. Polyakov, Phys. Lett. {\bf B428} (1998) 105,
hep-th/9802109; E. Witten, ATMP {\bf 2} (1998) 253,
hep-th/9802150.} \lref\albion{V. Balasubramanian, P. Kraus and A.
Lawrence, Phys. Rev. {\bf D59} (1999) 046003, hep-th/9805171.}
\lref\dimo{E. Kolb and M. Turner, {\it The Early Universe},
Addison-Wesley, Redwood City, 1990.} \lref\spesta{ D.N. Spergel
and P.J. Steinhardt, Phys. Rev. Lett. {\bf 84} (2000) 3760,
astro-ph/9909386.} \lref\peebles{P.J.E. Peebles,
astro-ph/0002495.} \lref\dine{M. Dine and A. Nelson, Phys. Rev.
{\bf D48} (1993) 1277, hep-th/9303230.}

\lref\cen{R. Cen, astro-ph/0005206.}
\lref\ygmn{Y. Grossman and M.
Neubert, Phys. Lett. {\bf B474} (2000) 361, hep-ph/9912408}
\lref\joe{ J.~Dai, R.~G.~Leigh and J.~Polchinski,
Mod.\ Phys.\ Lett.\  {\bf A4}, 2073 (1989); P.~Horava,
Phys.\ Lett.\  {\bf B231}, 251 (1989); P.~Horava,
Nucl.\ Phys.\  {\bf B327}, 461 (1989); G.~Pradisi and A.~Sagnotti,
Phys.\ Lett.\  {\bf B216}, 59 (1989); J.~Polchinski,
Phys.\ Rev.\ Lett.\ {\bf 75}, 4724 (1995), hep-th/9510017. }
\lref\dfgk{O. DeWolfe, D. Freedman, S. Gubser and A. Karch,
``Modeling the Fifth-Dimension with Scalars and Gravity,'' Phys.
Rev. {\bf D62} (2000) 046008.}
\lref\add{N. Arkani-Hamed, S.
Dimopoulos and G. Dvali, Phys. Lett. {\bf B429} (1998) 263,
hep-th/9803315; I. Antoniadis, N. Arkani-Hamed, S. Dimopoulos and
G. Dvali, Phys. Lett. {\bf B436} (1998) 257, hep-th/9804398; N.
Arkani-Hamed, S. Dimopoulos and G. Dvali, Phys. Rev. {\bf D59}
(1999) 086004, hep-th/9807344.}
\lref\dkklsII{S. Dimopoulos, S.
Kachru, N. Kaloper, A. Lawrence and E. Silverstein, preprint to
appear.}
\lref\shiu{G. Shiu and S.H. Tye, ``TeV Scale Superstring
and Extra Dimensions,'' Phys. Rev. {\bf D58} (1998) 106007,
hep-th/9805157.}
\lref\hw{P. Ho\v rava and E. Witten, Nucl. Phys.
{\bf B460} (1996) 506, hep-th/9510209.}
\lref\edunify{E. Witten, Nucl. Phys. {\bf B471} (1996) 135,
hep-th/9602070.}
\lref\svw{S. Sethi, C. Vafa and E. Witten,
``Constraints on Low Dimensional String Compactifications,'' Nucl.
Phys. {\bf B480} (1996) 213, hep-th/9606122.}

\lref\sutter{S. Dimopoulos and D. Sutter, Nucl. Phys. {\bf B452} (1995) 496.}

\lref\ovrut{For a review with references, see: A. Lukas, B. Ovrut
and D. Waldram, ``Heterotic M-theory Vacua with Five-branes,''
Fortsch. Phys. {\bf 48} (2000) 167, hep-th/9903144.}

\lref\braneworld{Recent attempts to construct such models can be
found in: G. Aldazabal, S. Franco, L.E. Ibanez, R. Rabadan and
A.M. Uranga, ``D=4 Chiral String Compactifications from
Intersecting Branes,'' hep-th/0011073; G. Aldazabal, S. Franco,
L.E. Ibanez, R. Rabadan and A.M. Uranga, ``Intersecting Brane
Worlds,'' hep-ph/0011132; G. Aldazabal, L.E. Ibanez, F. Quevedo
and A.M. Uranga, ``D-branes at Singularities: A Bottom Up Approach
to the String Embedding of the Standard Model,'' JHEP {\bf 0008}
(2000) 002, hep-th/0005067.}

\lref\adsbreak{S. Kachru and E. Silverstein, ``4d Conformal Field
Theories and Strings on Orbifolds,'' Phys. Rev. Lett. {\bf 80}
(1998) 4855, hep-th/9802183\semi A. Lawrence, N. Nekrasov and C.
Vafa, ``On Conformal Field Theories in Four Dimensions,'' Nucl.
Phys. {\bf B533} (1998) 199, hep-th/9803015\semi J. Distler and F.
Zamora, ``Non-Supersymmetric Conformal Field Theories from Stable
Anti-de Sitter Spaces,'' Adv. Theor. Math. Ph. {\bf 2} (1999)
1405, hep-th/9810206.}
\lref\mart{D.~E.~Kaplan, G.~D.~Kribs and M.~Schmaltz,
Phys.\ Rev.\ {\bf D62}, 035010 (2000);
Z.~Chacko, M.~A.~Luty, A.~E.~Nelson and E.~Ponton,
JHEP {\bf 0001}, 003 (2000);
M.~Schmaltz and W.~Skiba,
Phys.\ Rev.\ {\bf D62}, 095005 (2000);
Phys.\ Rev.\  {\bf D62}, 095004 (2000);
D.~E.~Kaplan and G.~D.~Kribs,
JHEP {\bf 0009}, 048 (2000).}
\lref\bklt{V. Balasubramanian, P. Kraus, A. Lawrence and S.
Trivedi, ``Holographic probes of anti-de Sitter spacetimes,''
Phys. Rev. {\bf D59}\ (1999) 104021; hep-th/9808017.} \lref\bgl{V.
Balasubramanian, S.B. Giddings and A. Lawrence, ``What do CFTs
tell us about anti-de Sitter spacetimes?'' JHEP {\bf 9903}\ (1999)
001; hep-th/9902052.} \lref\bdhm{T. Banks, M.R. Douglas, G.T.
Horowitz and E. Martinec, ``AdS dynamics from conformal field
theory,'' hep-th/9808016.} \lref\scatt{I.R. Klebanov, ``World
volume approach to absorption by non-dilatonic branes,'' Nuc.
Phys. {\bf B496}\ (1997) 231.} \lref\brfreed{P. Breitenlohner and
D.Z. Freedman, ``Positive energy in anti-de Sitter backgrounds and
gauged extended supergravity,'' Phys. Lett. {\bf B115}\ (1982)
197; ibid., ``Stability in gauged extended supergravity,'' Ann.
Phys. NY {\bf 144}\ (1982) 249; L. Mezinescu and P.K. Townsend,
``Stability at a local maximum in higher-dimensional anti-de
Sitter space and applications to supergravity,'' Ann. Phys. NY
{\bf 160}\ (1985) 406.} \lref\Igor{S. Gubser, I. Klebanov, and A.
Peet, ``Entropy and Temperature of Black 3-Branes'', Phys. Rev.
{\bf D54} (1996) 3915, hep-th/9602135.} \lref\IgorII{I. Klebanov,
``Wordvolume Approach to Absorption by Nondilatonic Branes'',
Nucl. Phys. {\bf B496} (1997) 231; S. Gubser, I. Klebanov, A.
Tseytlin, ``String Theory and Classical Absorption by
Three-Branes'', Nucl. Phys. {\bf B499} (1997) 217.}
\lref\gravbox{J. Lykken, R. Myers, J. Wang, ``Gravity in a Box'',
JHEP {\bf 0009} (2000) 009.} \lref\mirpe{E.A. Mirabelli and M.E.
Peskin, Phys. Rev. {\bf D58} (1998) 065002, hep-th/9712214.}

\lref\appsym{N. Arkani-Hamed and S. Dimopoulos, hep-ph/9811353; N.
Arkani-Hamed, S. Dimopoulos, G. Dvali and J. March-Russell,
hep-ph/9811448; N. Arkani-Hamed and M. Schmaltz, Phys. Rev. {\bf
D61} (2000) 033005, hep-ph/9903417; N. Arkani-Hamed, Y. Grossman
and M. Schmaltz, Phys. Rev. {\bf D61} (2000) 115004,
hep-ph/9909411.}
\lref\selftuning{E. Verlinde and H. Verlinde, JHEP {\bf 0005} (2000) 034;
N. Arkani-Hamed, S. Dimopoulos,
N. Kaloper and R. Sundrum, Phys. Lett. {\bf B480} (2000) 193,
hep-th/0001197; S. Kachru, M. Schulz and E. Silverstein, Phys.
Rev. {\bf D62} (2000) 045021, hep-th/0001206; {it ibid.}, Phys.
Rev. {\bf D62} (2000) 085003;
R. Bousso  and J. Polchinski, JHEP {\bf 0006} (2000) 006;
J.L. Feng, J. March-Russell, S. Sethi and F. Wilczek, hep-th/0005276. }
\lref\heterotic{D.J. Gross, J.A.
Harvey, E. Martinec and R. Rohm, Phys. Rev. Lett {\bf 54} (1985)
502; Nucl. Phys. {\bf B256} (1985) 253; Nucl. Phys. {\bf B267}
(1986) 75.}

\lref\RSAdS{J. Maldacena, unpublished;
E. Witten, unpublished;  S. Gubser, ``AdS/CFT
and Gravity,'' hep-th/9912001 ; S. Giddings, E. Katz and L. Randall,
``Linearized Gravity in Brane Backgrounds,''
JHEP {\bf 0003} (2000) 023, hep-th/0002091;
M. Duff and J. Liu, ``On the Equivalence of the Maldacena
and Randall-Sundrum Pictures,'' Phys. Rev. Lett. {\bf 85} (2000) 2052,
hep-th/0003237; S. Giddings and E. Katz, ``Effective Theories and
Black Hole Production in Warped Compactifications,'' hep-th/0009176;
N. Arkani-Hamed, M. Porrati and L. Randall, ``Holography and
Phenomenology,'' hep-th/0012148; R. Rattazzi and A. Zaffaroni,
``Comments on the Holographic Picture of the Randall-Sundrum Model,''
hep-th/0012248.
}
\lref\juan{J. Maldacena, ``The Large N Limit of Superconformal
Field Theories and Supergravity,'' ATMP {\bf 2} (1998) 231, hep-th/9711200.}
\lref\RSII{L. Randall and R. Sundrum, ``An Alternative to
Compactification,'' Phys. Rev. Lett. {\bf 83} (1999) 4690, hep-th/9906064.}
\lref\RSI{L. Randall and R. Sundrum, ``A Large Mass Hierarchy from
a Small Extra Dimension,'' Phys. Rev. Lett. {\bf 83} (1999) 3370,
hep-th/9905221.}
\lref\AdSCFT{S. Gubser, I. Klebanov and A. Polyakov, ``Gauge Theory
Correlators from Noncritical String Theory,'' Phys. Lett. {\bf B428}
(1998) 105, hep-th/9802109; E. Witten, ``Anti-de Sitter Space and
Holography,'' ATMP {\bf 2} (1998) 253, hep-th/9802150.}
\lref\albion{V. Balasubramanian, P. Kraus and A. Lawrence, ``Bulk vs
Boundary Dynamics in Anti-de Sitter Space-Time,'' Phys. Rev.
{\bf D59} (1999) 046003, hep-th/9805171.}
\lref\dimo{S. Dimopoulos, ``LHC, SSC and the Universe'', Phys. Lett.
{\bf B246} (1990) 347; N. Arkani-Hamed, L.J. Hall, C. Kolda
and H. Murayama, ``A New Perspective on Cosmic Coincidence Problems",
Phys. Rev. Lett. {\bf 85} (2000) 4434, astro-ph/0005111.}
\lref\spesta{ D.N. Spergel and P.J. Steinhardt,
``Observational Evidence for Selfinteracting Cold Dark Matter",
Phys. Rev. Lett. {\bf 84} (2000) 3760, astro-ph/9909386.}
\lref\peebles{P.J.E. Peebles , ``Fluid Dark Matter", astro-ph/0002495.}
\lref\cen{Renyue Cen, ``Decaying Cold Dark Matter Model and Small-Scale
Power",  astro-ph/0005206}
\lref\addv{
I. Antoniadis, S. Dimopoulos and G. Dvali, ``Millimeter Range Forces in Superstring Theories
with Weak Scale Compactification", Nucl. Phys. {\bf B516} (1998) 70, hep-ph/9710204 }

\lref\smallone{S. Dimopoulos, S. Kachru, N. Kaloper, A. Lawrence
and E. Silverstein, ``Small Numbers From Tunneling Between Brane
Throats,'' hep-th/0104239.}

\lref\adqp{
I. Antoniadis, S. Dimopoulos, A. Pomarol and M. Quiros, ``Soft Masses in
Theories with Supersymmetry
Breaking by TeV Compactification", Nucl. Phys. {\bf B544} (1999) 503,
hep-ph/9810410 }
\lref\mirpe{E.A. Mirabelli and M.E. Peskin, ``Transmission of Supersymmetry
Breaking from a Four-Dimensional Boundary", Phys. Rev. {\bf D58} (1998) 065002,
hep-th/9712214 }
\lref\ygmn{Y. Grossman and M. Neubert, ``Neutrino Masses and Mixings
in Nonfactorizable Geometry", Phys. Lett. {\bf B474} (2000) 361, hep-ph/9912408}
\lref\gherpom{T. Gherghetta and A. Pomarol, ``A Warped Supersymmetric
Standard Model,'' hep-ph/0012378.}
\lref\GKP{S. Giddings, S. Kachru and J. Polchinski, ``Hierarchies from
Fluxes in String Compactifications,'' to appear.}
\lref\joereview{J. Polchinski, ``TASI Lectures on D-branes,'' hep-th/9611050.}
\lref\dfgk{O. DeWolfe, D. Freedman, S. Gubser and A. Karch, ``Modeling
the Fifth-Dimension with Scalars and Gravity,'' Phys. Rev. {\bf D62}
(2000) 046008.}
\lref\led{N. Arkani-Hamed, S. Dimopoulos and G. Dvali, ``The
Hierarchy Problem and New Dimensions at a Millimeter,'' Phys. Lett.
{\bf B429} (1998) 263, hep-th/9803315; I. Antoniadis, N. Arkani-Hamed,
S. Dimopoulos and G. Dvali, ``New Dimensions at a Millimeter to a Fermi
and Superstrings at a TeV,''
Phys. Lett. {\bf B436} (1998) 257, hep-th/9804398.}
\lref\hw{P. Ho\v rava and E. Witten, ``Heterotic and Type I String Dynamics
from Eleven-Dimensions,'' Nucl. Phys. {\bf B460} (1996) 506, hep-th/9510209.}
\lref\HV{H. Verlinde, ``Holography and Compactification,'' Nucl. Phys.
{\bf B580} (2000) 264, hep-th/9906182; S. Giddings,
S. Kachru and J. Polchinski, ``Hierarchies from Fluxes in
String Compactifications",  hep-th/0105097.}
\lref\edunify{E. Witten, ``Strong Coupling Expansion of Calabi-Yau
Compactification,'' Nucl. Phys. {\bf B471} (1996) 135, hep-th/9602070.}
\lref\svw{S. Sethi, C. Vafa and E. Witten, ``Constraints on Low
Dimensional String Compactifications,'' Nucl. Phys. {\bf B480} (1996)
213, hep-th/9606122.}

\lref\braneworld{Recent attempts to construct such models can be found in:
G. Aldazabal, S. Franco, L.E. Ibanez, R. Rabadan and A.M. Uranga,
``D=4 Chiral String Compactifications from Intersecting Branes,''
hep-th/0011073; G. Aldazabal, S. Franco, L.E. Ibanez, R. Rabadan
and A.M. Uranga, ``Intersecting Brane Worlds,'' hep-ph/0011132;
G. Aldazabal, L.E. Ibanez, F. Quevedo and A.M. Uranga, ``D-branes
at Singularities: A Bottom Up Approach to the String Embedding of
the Standard Model,'' JHEP {\bf 0008} (2000) 002, hep-th/0005067.}

\lref\adsbreak{S. Kachru and E. Silverstein, ``4d Conformal Field
Theories and Strings on Orbifolds,'' Phys. Rev. Lett. {\bf 80}
(1998) 4855, hep-th/9802183\semi
A. Lawrence, N. Nekrasov and C. Vafa, ``On Conformal Field Theories in
Four Dimensions,'' Nucl. Phys. {\bf B533} (1998) 199, hep-th/9803015\semi
J. Distler and F. Zamora, ``Non-Supersymmetric Conformal Field Theories
from Stable Anti-de Sitter Spaces,'' Adv. Theor. Math. Ph. {\bf 2} (1999)
1405, hep-th/9810206.}

\lref\suGUT{ E. Cremmer, S. Ferrara, L. Girardello, and A. Van
Proyen, Phys. Lett. {\bf B116} (1982) 231;
A. Chamseddine, R. Arnowitt, and P. Nath, Phys. Rev. Lett.
{\bf 49} (1982) 970;
R. Barbieri, S. Ferrara, and C. Savoy, Phys. Lett. {\bf B110} (1982)
343;
L. J. Hall, J. Lykken, and S. Weinberg, Phys. Rev. {\bf D27}(1983)
2359.}

\lref\nil{H.P. Nilles, Phys. Lett. {\bf B115} (1982) 193;
Nucl. Phys. {\bf B217} (1983) 366.}
\lref\kaplu{V.S. Kaplunovsky and J. Louis, Phys. Lett. {\bf B306} (1993) 269.}

\lref\gauco{M. Dine, R. Rohm, N. Seiberg, E. Witten, Phys. Lett. {\bf
B156} (1985) 55.}

\lref\bklt{V. Balasubramanian, P. Kraus, A. Lawrence and S. Trivedi,
``Holographic probes of anti-de Sitter spacetimes,''
Phys. Rev. {\bf D59}\ (1999) 104021;
hep-th/9808017.}
\lref\bgl{V. Balasubramanian, S.B. Giddings and A. Lawrence,
``What do CFTs tell us about anti-de Sitter spacetimes?''
JHEP {\bf 9903}\ (1999) 001; hep-th/9902052.}
\lref\bdhm{T. Banks, M.R. Douglas, G.T. Horowitz and E. Martinec,
``AdS dynamics from conformal field theory,'' hep-th/9808016.}
\lref\scatt{I.R. Klebanov, ``World volume approach to
absorption by non-dilatonic branes,'' Nuc. Phys.
{\bf B496}\ (1997) 231.}
\lref\brfreed{P. Breitenlohner and D.Z. Freedman,
``Positive energy in anti-de Sitter backgrounds
and gauged extended supergravity,'' Phys. Lett.
{\bf B115}\ (1982) 197; ibid.,
``Stability in gauged extended supergravity,'' Ann. Phys. NY
{\bf 144}\ (1982) 249; L. Mezinescu and P.K. Townsend,
``Stability at a local maximum in higher-dimensional
anti-de Sitter space and applications to supergravity,''
Ann. Phys. NY {\bf 160}\ (1985) 406.}
\lref\Igor{S. Gubser, I. Klebanov, and A. Peet, ``Entropy and
Temperature of Black 3-Branes'', Phys. Rev. {\bf D54} (1996) 3915,
hep-th/9602135.}
\lref\IgorII{I. Klebanov, ``Wordvolume Approach to Absorption
by Nondilatonic Branes'', Nucl. Phys. {\bf B496} (1997) 231;
S. Gubser, I. Klebanov, A. Tseytlin, ``String Theory and Classical
Absorption by Three-Branes'', Nucl. Phys. {\bf B499} (1997) 217.}
\lref\gravbox{J. Lykken, R. Myers, J. Wang, ``Gravity in a Box'',
JHEP {\bf 0009} (2000) 009.}
\lref\gmam{}

\lref\wessbagg{J. Wess and J. Bagger, {\it Supersymmetry and
Supergravity}, 2nd. ed., Princeton University Press (1992).}

\def\ap{\alpha'}
\def\ie{{\it i.e.}}
\def\CO{{\cal O}}
\def\CB{{\cal B}}
\def\CN{{\cal N}}
\def\luv{\Lambda_{UV}}
\def\frac#1#2{{#1 \over #2}}
\def\p{\partial}
\def\ie{{\it i.e.}}
\def\cf{{\it c.f.}}

\def\tp{{\darr\p}}

\Title{\vbox{\baselineskip12pt\hbox{hep-th/0106128}
\hbox{SLAC-PUB-8862}\hbox{SU-ITP-01/22}
\hbox{NSF-ITP-01-61}
\hbox{DUKE-CGTP-01-06}}}
{\vbox{
\centerline{Generating Small Numbers by Tunneling}
\bigskip
\centerline{in Multi-Throat Compactifications}
}}
\bigskip
\centerline{Savas Dimopoulos$^{1}$, Shamit Kachru$^{1,2}$,
Nemanja Kaloper$^{1,2}$,}
\centerline{Albion Lawrence$^{1,3}$, and Eva Silverstein$^{1,2}$}
\bigskip
\centerline{$^{1}${\it Department of Physics and SLAC,
Stanford University, Stanford, CA 94305/94309}}
\centerline{$^{2}${\it Institute for Theoretical Physics,
University of California, Santa Barbara, CA 93106}}
\centerline{$^{3}${\it Center for Geometry and Theoretical Physics,
Duke University, Durham, NC 27708}}

\bigskip
\bigskip
\noindent

A generic F-theory compactification containing many D3 branes
develops multiple brane throats. The interaction of observers
residing inside different throats involves tunneling
suppression and, as a result, is very weak. This suggests a
new mechanism for generating small numbers in Nature. One
application is to the hierarchy problem: large
supersymmetry breaking near the unification scale inside a
shallow throat causes TeV-scale SUSY-breaking inside the
standard-model throat. Another application, inspired
by nuclear-decay, is in designing naturally long-lived
particles: a cold dark matter particle residing near the
standard model brane decays to an approximate CFT-state of
a longer throat within a Hubble time. This suggests that
most of the mass of the universe today could consist of CFT-matter
and may soften structure formation at sub-galactic scales.
The tunneling calculation demonstrates that the coupling
between two throats is dominated by higher dimensional modes
and consequently is much larger than a naive application of
holography might suggest.

\bigskip
\Date{June 2001}

\newsec{Introduction}

The enormous differences in scales that appear in Nature present a
formidable challenge for any unified theory of forces.  Grand
unification addresses this problem by postulating an energy desert
separating the gravitational and the electroweak scale \ggl. The
supersymmetric version of this picture  \dg, the supersymmetric
standard model (SSM), has had a quantitative success: the
unification prediction of the value of the weak mixing angle \dg,
subsequently confirmed by the LEP and SLC experiments.  While this
picture is attractive, it leaves many fundamental questions
unanswered. There are 125 unexplained parameters, many of them
mysteriously small; these include the masses of the three
generations of particles and the cosmological constant. String
theory provides a natural framework for addressing these
questions. Many scenarios for string phenomenology involve {\it
localized} gauge fields.  Perhaps the simplest is the minimal
Ho\v{r}ava-Witten theory \refs{\hw,\edunify}; other models use
``D-brane'' defects on which gauge dynamics occurs \joe. A
striking possibility emerging from these ingredients is a new
explanation the weakness of gravity \refs{\add}. These ideas are
providing new avenues for exploring physics beyond the Standard
Model, and novel mechanisms for explaining small numbers
\refs{\appsym,\RSI,\selftuning}.

Ho\v{r}ava-Witten theory and the perturbative $E_8 \times
E_8$ heterotic string \heterotic\ have been well studied
in calculable, weakly-coupled
regimes. In this note we will study string phenomenology in a
different calculable regime, which can arise when there are many
branes transverse to the compactification manifold $M$. The
tension of the branes curves the space around them. The
backreaction is proportional to the sum of brane tensions, and
therefore to the total number of branes in some region of space.
Hence solitary branes have little effect and their neighborhood is
nearly flat. Such ``dilute gases'' of branes are commonly studied
in e.g. perturbative string orientifold constructions. In other
regimes of couplings where a (super)gravity description is valid,
large stacks of branes in the compactification manifold $M$
significantly alter the metric on $M$. The regions of space where
the branes reside may be viewed as gravitational funnels, or
throats. Examples in this regime arise in F-theory
compactifications on elliptic Calabi-Yau fourfolds \HV.  From the
4d point of view the geometry is ``warped'' -- the scale factor of
the 4d metric depends on the distance down the throat.
\ifig\octopus{The Calabi-Yau octopus.  $N_i$ is the number of
branes in a given region.} {\epsfxsize3.0in\epsfbox{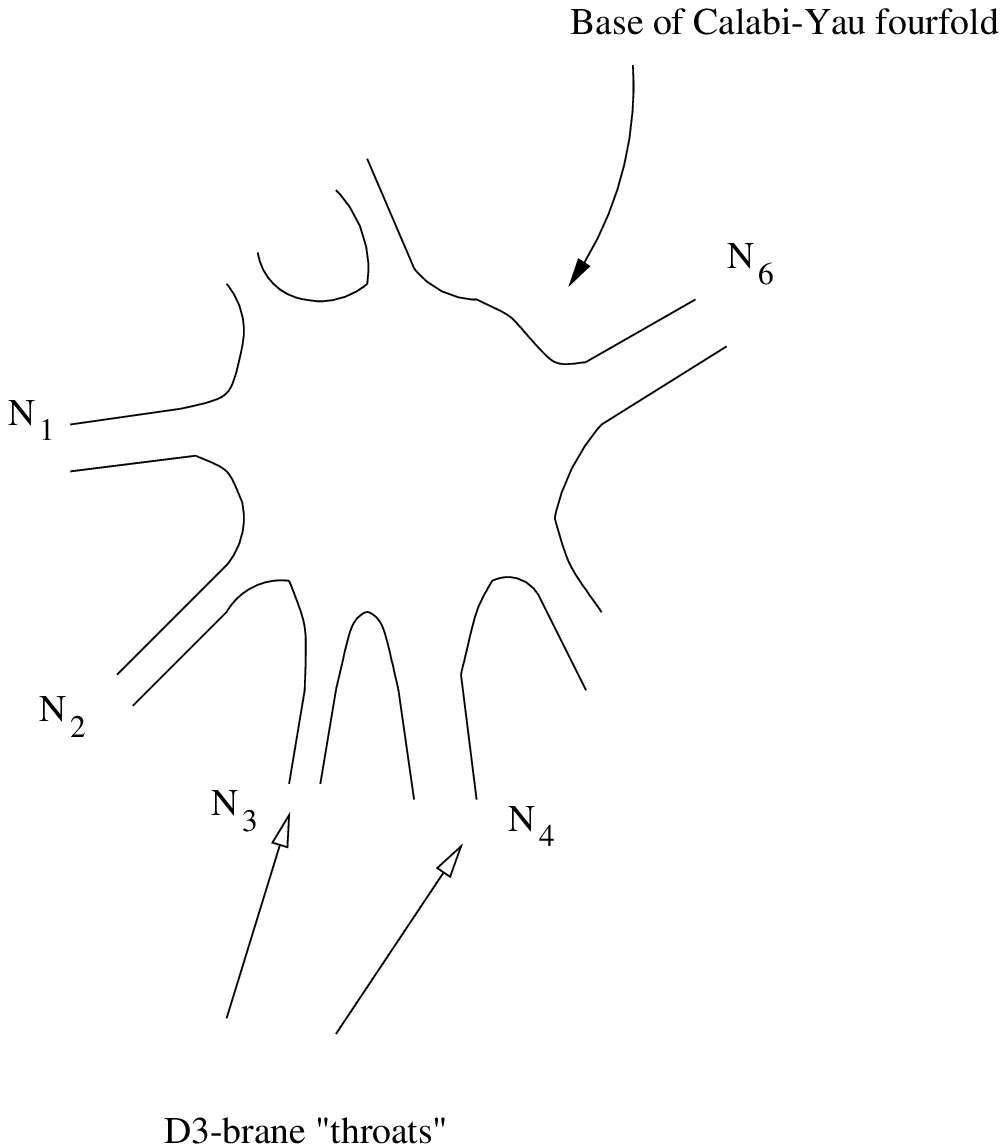}}

One class of such scenarios was suggested by H. Verlinde \HV,
who noted that a wide class of F-theory backgrounds have this property.
In F-theory on a Calabi-Yau fourfold $X$, tadpole cancellation accommodates
the introduction of
\eqn\dthree{N = {\chi(X)\over 24}}
D3 branes transverse to $X$ \svw.
$N >> 1$ is readily achieved for simple choices of $X$.
In a suitable limit of couplings, the correct F-theory
geometry will involve D3-brane throats
glued into the base $\CB$ of the elliptically fibered fourfold.
A generic possibility is that there are several
such throats glued into $\CB$;
these correspond to different stacks of branes separated on
$\CB$. The ensuing geometry of the compactification
resembles an ``octopus,''
where the legs represent throats arising from brane stacks, as
depicted in Fig. 1. The (super)gravity modes in the throat and the
low-energy field theory on the branes are dual to each other
\juan: the degrees of freedom localized at the ends of the throats
are dual to infrared (IR) excitations of the field theory, while
the excitations closer to the mouth of the throat are dual to the
ultraviolet (UV) degrees of freedom.

This geometry suggests a new mechanism for generating small
numbers in 4d physics. The mutual couplings of the IR degrees of
freedom residing in different throats are suppressed, as these
modes must tunnel through the bulk to communicate.
This leads to small couplings between
degrees of freedom localized in different throats.

We make this intuition precise in a 5d toy model of Fig. 1, which
appears in Fig. 2. The detailed form of the metric in the throats
does ${\it not}$ play an important role in our mechanism.  The
small numbers do not arise
from exponential redshifting associated with long throats,
but rather from the tunneling suppression emerging due to
an effective potential barrier for localized
KK modes. Thus the effects
we study would persist with generic warped metrics (including
those with only power-law warping).
However for simplicity and to facilitate a holographic interpretation we will
take AdS metrics in the throats of our toy model\foot{Another option,
which would not change
the generic physics of the results, would be to use
the multi-center D3-brane solution and generalize the
computations of Klebanov et al \refs{\Igor,\IgorII} to this case
of tunneling between threebrane throats.  An intermediate
option would be to introduce a flat space between the
AdS slices in Figure 2, as in \gravbox.  Here we will stick
with the simplest case depicted in Figure 2.}.
We will join these brane throats
by a ``UV brane'' playing the role of the bulk of $M$, choosing
the scale of physics
at the UV brane to be $1/L$ for simplicity. In general
one could choose this scale $M_{UV}$ to be unrelated to either
$L$ or $M_4$, reflecting the freedom to vary the geometrical moduli
near the stacks of branes in the microscopic Calabi-Yau picture.
The AdS throats terminate at infrared branes on either side.
Such ``end-of-the-world" branes mimic a mass gap or be IR-freedom
of a generic low
energy theory below some scale $\Lambda_{IR}$.
We can use the scale-radius duality to cut off the gravitational background
at a radius dual to $\Lambda_{IR}$.

We
then show explicitly that the KK modes in
adjacent throats must tunnel to communicate between
different throats, and that this
effect generates small numbers.

\ifig\twoface{A cartoon of two throats.
The five-dimensional wave equation, for fixed
four-momentum, reduces to a one-dimensional
Schr\"odinger equation with a potential
barrier as shown.  When the throat
geometry is anti-de Sitter, the potential
is simply the warp factor in conformal coordinates.}
{\epsfxsize3.0in\epsfbox{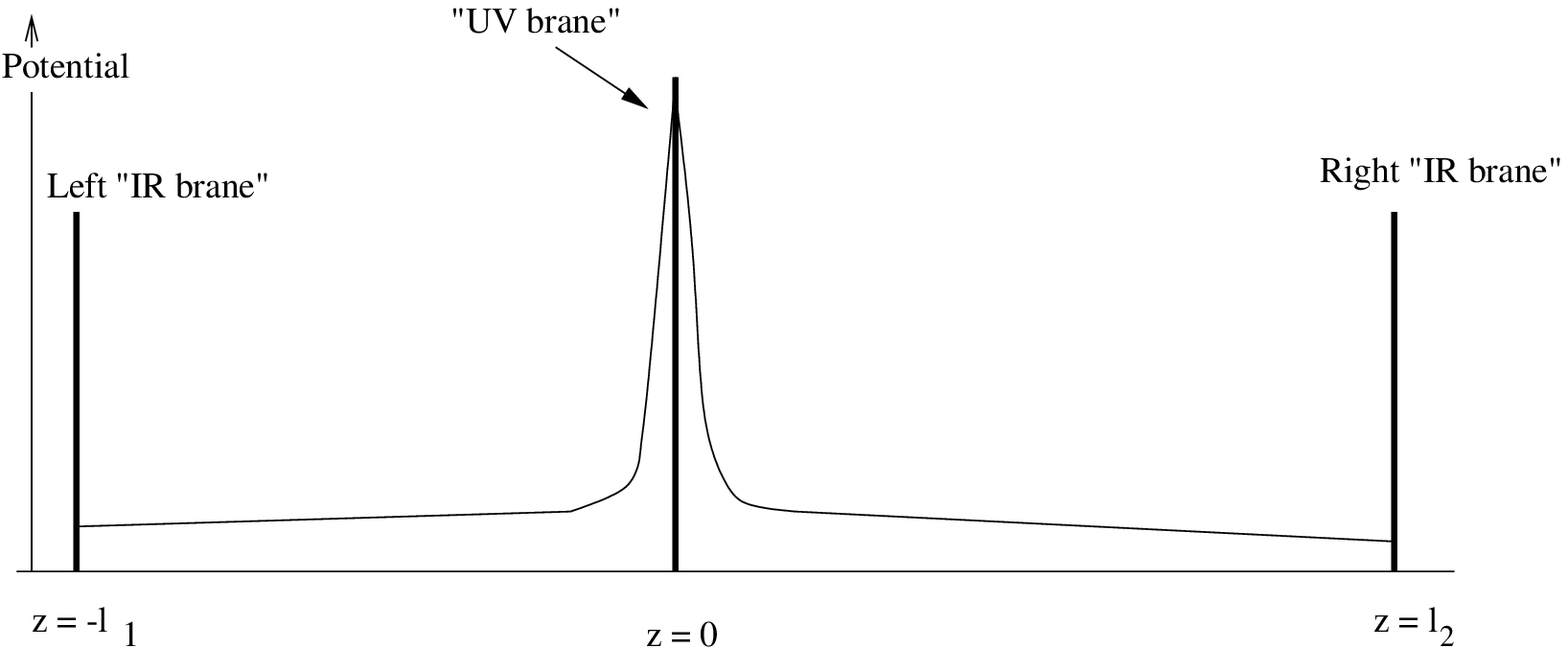}}


To gain the insight in the physics of this mechanism
one should understand it from the point of view of an effective 4d
theory. To deduce it, one recalls that the string dynamics far down each throat is
dual to a strongly coupled field theory.
For AdS warping, a natural guess for any single throat is that the low energy
effective theory is a conformal field theory
coupled to 4d gravity, with a Lagrangian of the form
\eqn\leff{
    {\cal L} ~=~ {\cal L}_{CFT} + \sum_{n>0} ({1\over M_4})^n
{\cal O}_n\ ;
}
here $M_4$ is the 4d Planck scale and
the ${\cal O}_n$ are irrelevant operators of dimension $4+n$.
In the two throat case, one might naively expect the effective
theory to be two
conformal field theories coupled to each other and to
gravity, with an effective Lagrangian of the form
\eqn\lefftwo{
{\cal L}_{tot} =
{\cal L}_{CFT 1}(\phi) + {\cal L}_{CFT 2}(\psi) +
\sum_{n>0} ({1\over M_4})^n {\cal O}_{n}(\phi,\psi)\ ;
}
here $\phi$ and $\psi$ represent the fields in the two CFTs.

However, in
the  higher-dimensional gravitational theory the
communication between the
throats arises from tunneling of the KK modes, and
they obey an effective Schr\"odinger equation which
only depends only on the background geometry in the two
throats, and not on the 4d Planck scale.  So in \leff,\lefftwo,
$M_4$ should be replaced by a scale
$M_{UV}$ arising from the background geometry.
There is a large regime where the geometrical moduli of
the compactification are chosen so that $M_{UV} \ll M_4$, so communication
between distinct throats can be much larger
than is suggested by \lefftwo, as is clear from the Calabi-Yau
picture of Figure 1.
To see this, one can take a double scaling
limit where the volume of $\CB$ diverges, while
the throats are kept at a fixed relative distance.
While $M_4 \to \infty$,
the communication between the throats persists.
This shows that in the 4d effective theory,
$M_4$ arises due to contributions from both the bulk of ${\cal B}$
and the throats, as
\eqn\rel{
    M_4^2 = M_5^3 L + M_{4,0}^2\ .
}
Thus we can take $M_{4,0}$ to be large while fixing
$M_5$ and $L$.

Further we find that
CFT modes generically mix with 10d or 5d
gravity modes.  In the strict near-horizon limit of
\juan, such modes are non-normalizable and non-fluctuating
due to the infinite volume of AdS space \refs{\AdSCFT,\albion}.
In the finite volume setup of Fig. 1, they become normalizable
and fluctuating.  So the operators ${\cal O}_n$ are not just
made up of operators from the original CFT.
We will find that this significantly enhances
some phenomenologically interesting processes. This
highlights new aspects of
the relationship between the physics
of RS scenarios and the AdS/CFT correspondence.

The organization of this paper is as follows.
In section 2, we will directly compute the amplitude for an excitation
in the left throat to tunnel into the right throat.
In section 3, we discuss a potential
application of this process in designing candidates for the
astrophysical
dark matter.
In section 4, we discuss the spectrum of 4d excitations in
our system and in a closely related system with supersymmetry
broken in one of the two throats.  We begin an exploration of
phenomenological aspects of this ``tunneling-mediation'' of
supersymmetry breaking.
In section 5 we present a discussion of the phenomenological aspects of
tunneling mediation. We discuss the
holographic 4d field
theory interpretation of our 5d gravity calculations in section 6.
A companion paper, which summarizes the essential physics of this
paper but with detailed calculations omitted, appears in \smallone.

\newsec{Tunneling and Glueball Decay}

Our generic compactification (Figs 1, 2) has a four-dimensional
description involving multiple coupled gauge theory sectors.
The gauge-invariant
excitations in each gauge theory sector will mix to some extent as
long as the interactions respect the symmetries of the
problem.  By the AdS/CFT correspondence, the gauge theory degrees
of freedom in the different throats are all comprised of the
same bulk degrees of freedom of
M theory.  At the level of the low-energy gravity modes, this
rephrases the question about the mixing of the two gauge
theory sectors into a relatively simple double-well potential
problem which we now analyze in detail.

For our explicit calculations we will work with the schematic
double well background of Figure 2.  The metric is given by
\eqn\metricL{ ds^2={L^2\over{(|z|+L)^2}}(\eta_{\mu\nu}dx^\mu
dx^\nu +dz^2), ~~~~ -l_1\le z\le l_2 } in conformally
flat coordinates.  The proper distances
between the UV branes and left and right branes respectively
are $R_{1,2}$ where $l_k=Le^{R_k/L}$. We will assume $l_1 \ge l_2$
throughout the remainder.
It will be useful to consider the limit where the UV brane is removed,
and 4d gravity is decoupled.
This is readily obtained if we replace $L$ in the denominator of the warp factor
in \metricL\ by $\zeta L$ and take the limit of the dimensionless
parameter $\zeta \rightarrow 0$, removing the cutoff in the $AdS$ space.

The analysis of the KK spectrum is a straightforward
generalization of that in \RSI.
Decompose the transverse traceless mode of the 5d graviton as:
$$
h_{\mu\nu}(x,z)= \sqrt{{L \over |z|+L}} e^{ip\cdot x}\psi_{\mu\nu}(z)\ ,
$$
and drop the $\mu\nu$ indices.
The equation of motion for $\psi$ is:
\eqn\eomL{
\psi''+(m^2-{{15}\over{4(|z|+L)^2}})\psi=0, ~~~~-l_1\le z\le l_2\ ,}
where $m^2 = p^2$ is the mass of the $4D$ KK mode.
$M_{4,5}$ do not appear in this effective
Schr\"odinger equation.
The solutions are given in terms of Bessel and Neumann functions
in each AdS throat:
\eqn\Lsoln{\psi(z)=\sqrt{m(|z|+L)}\biggl(
AJ_2[m(|z|+L)]+BN_2[m(|z|+L)]\biggr), }
where $A$ and $B$ are constants.

To compute the tunneling probability through the barrier to the right side,
we should impose the proper tunneling boundary conditions.
This means that for $z > 0$ the wavefunction $\psi$ will be purely outgoing, propagating
toward larger $z$, asymptotically approaching $e^{iz/L}$ for large $z$.
Since $J_2 (x)$ and $N_2 (x)$
behave as $\sqrt{2\over{\pi x}}\cos (x-{5\over 4}\pi)$ and
$\sqrt{2\over{\pi x}}\sin (x-{5\over 4}\pi)$ respectively for large argument $x$,
the purely outgoing wave to the right must be
the Hankel function $H_2^+(x)=J_2(x)+iN_2(x)$. On the left,
we will have both incoming and reflected waves, and so
the wavefunction is a linear combination of both Hankel
functions $H_2^\pm(x)$.
The solution takes the form:
\eqn\tunsol{\eqalign{
&\psi_1=\sqrt{m(|z|+L)}\biggl(AH_2^+(m(|z|+L))
+BH_2^-(m(|z|+L))\biggr), ~~~~z<0\cr &\psi_2=C\sqrt{m(|z|+L)}
H_2^+(m(|z|+L)), ~~~~z>0\cr } }

At the UV brane we must impose:
continuity of the wavefunction, \eqn\UVI{
\psi_1(0)=\psi_2(0)\ ; } and a matching condition,
which comes from perturbing the background geometry junction
condition on the UV brane:
\eqn\UVII{ \psi'_2(0)-\psi'_1(0)=-{3\over L}\psi(0). }
These imply:
\eqn\tunC{\eqalign{
&B={{H_2^+(mL)}\over{H_2^-(mL)}}(C-A)\cr
&C={{H_1^-(mL)H_2^+(mL)-H_1^+(mL)H_2^-(mL)}\over
{H_1^+(mL)H_2^-(mL)+H_1^-(mL)H_2^+(mL)}}A\ . } }

The transition probability is \eqn\Pgen{ P={{|\vec
j_{out}|}\over{|\vec j_{in}|}} } where $\vec j_i$ are the
currents obtained from the wavefunctions \tunsol, \tunC.
The $z$-component of the current
$j_z=-i(\psi^*\psi'-\psi\psi^{*'})$ is constant;
we can evaluate it at large $|z|$, where the
Hankel functions turn into Fourier modes.  The
tunneling probability is: \eqn\probH{
P=\biggl|{{{H_1^-(mL)H_2^+(mL)-H_1^+(mL)H_2^-(mL)}\over
{2H_1^+(mL)H_2^+(mL)}} }\biggr|^2 }

When the left throat is terminated by an IR brane at $z=-l_1$,
there is an additional boundary condition,
$$\psi'_1 = {3 \over
2(l_1+L)} \psi_1 (l_1)\ .$$
Therefore the spectrum
of the left-localized states has a gap; the
mass $m$ of the tunneling glueball mode is of
order $m\sim n/l_1\sim {n\over L}e^{-R_1/L}$, where $n$ is an integer.
Since $mL\sim e^{-R_1/L}<<1$, we can expand the Hankel
functions appearing in our result \probH\ for small argument and find that:
\eqn\Pfin{ P\sim m^4L^4 \sim n^{4} e^{-4R_1/L} \ .}

Since the incoming flux goes like $1/l_1\sim m/n$, the
tunneling rate is given by \eqn\ratefin{ \Gamma\sim
{\pi^2\over{16 n}}L^4 m^5\sim {{\pi^2 n^4} \over{16}}{1\over L}e^{-5R_1/L}
} In four-dimensional terms, this is the
decay rate of a glueball from the left gauge theory sector into
glueballs from the right gauge theory sector.  This is important
for our discussion of dark matter in \S3.

\newsec{Dark Matter}

It is fascinating to contemplate the
possibility that the dark matter
which constitutes about 90 \% of the mass of the universe
is described by a CFT, and that we are immersed inside an
ocean of scale-free matter.\foot{This possibility has been
entertained independently by many physicists, including T. Banks, M. Dine,
and J. Maldacena.} In its simplest form this idea
is in conflict with observation: CFT matter would have a
relativistic equation of state, acting as hot dark matter
(HDM); but the large scale structure of the universe
suggests that non-relativistic, cold dark matter (CDM)
dominates the dynamics of the universe since about $t\sim
{10^4}$ years, when the temperature of the visible matter
was $\sim 10 eV $. A way to bypass this difficulty is to
postulate that the universe has, until recently, been
dominated by an unstable CDM particle which decays into CFT
matter, with a lifetime of order of the age of the universe.
This requires a coincidence: the CDM particle must have the
right abundance to be a good DM candidate, as well as the
right couplings to predominantly decay into CFT matter with
a lifetime about as long as the age of the universe. How
likely is this coincidence? To answer this we need a
concrete model of the CDM, the CFT matter, and their
couplings.

A simple model is suggested by the setup of the
previous section: Consider the left brane in figure 2 to be
at some finite scale, perhaps not far from the electroweak
scale $\sim TeV$, and eliminate the right brane by moving
it off to infinity. Imagine that the primordial CDM
particles are the lowest-energy, left-localized modes of
some five-dimensional field -- the
gravitons or some other bulk particles -- located next to the
left IR brane. These modes decay by tunneling into the CFT matter
residing in the right AdS throat. This setup is
sufficiently concrete that it allows a detailed calculation
of the abundance and the decay properties of the CDM. The
lifetime for the decay of CDM into CFT states is given by
Eq. \ratefin\ and depends on the mass of the particle and
the AdS radius.

We can use further data to constrain our model.
Several observations suggest that the universe has been
dominated, at least until recently, by CDM at the critical
density of $ \sim 10^{-29} gr/cm^{3}$. Under certain
conditions, it is possible to infer the necessary
microscopic properties that can lead to this abundance. For
example, particles that are in thermal and chemical
equilibrium --with zero chemical potential-- when the
universe is at a temperature of order of their mass have an
abundance determined by the non-equilibrium
dynamics of the expanding universe. Their leftover
abundance today can be calculated; it is the
result of their failure to totally
annihilate with their antiparticles because of the
expansion of the universe. The energy density,
for particles that are non-relativistic when they
stop annihilating, is:
\eqn\density{{\rho \over \rho_c} \simeq {1 \over 10^3 <\sigma v>}
{1 \over M_{Pl} \times 2.7 K} \simeq {1 \over 10^3 <\sigma v>}
{1 \over TeV^2}\ ,}
where we have used the fact that $TeV$ is the geometric mean between
the Planck scale and the millimeter, $TeV \simeq \sqrt{M_{Pl} \times 2.7K}$.
Here $\rho$ is the energy density of these particles at temperature $T$
(which we have set to the present day temperature),
$\rho_c$ is the critical energy density,
and $\sigma$ is the annihilation cross-section
of dark matter:
\eqn\acs{
\sigma \simeq {\alpha^2 \over m^2} N \ .}
Here $\alpha = g^2/4\pi$ is the coupling which dominates
annihilation, $m$ is the particle's mass, so that their
average speed is $v \sim 10^{-3} c$, and $N \sim 100$ is the
number of SM particles \dimo.
We see that a weakly interacting particle, such as the LSP,
must weigh about 100 GeV
to critically close the universe. However,  a strongly interacting
particle whose annihilation cross section saturates unitarity will
critically close the universe if its mass is about $ TeV$.

Introduce a bulk particle called the ``bulky'',
different than the graviton; let it carry a
``Bulk Parity'' or B-parity symmetry, under which
it changes sign. Decomposing the 5-d bulky into 4-d KK
modes, we find that the
lightest particle is the massless
bulky living deep inside the right throat. On the left
we have a quasi-stable
bulky which decays to right-bulkies with a lifetime of
order of \ratefin. Identify the left-brane with the
standard model brane on which there are $\sim 100$
particles. B-parity prevents the lightest left-bulky from
rapidly decaying into SM particles. On the other hand,
pairs of left-bulkies annihilate predominantly into
SM particles which reside close by in the fifth dimension. The
couplings responsible for bulky annihilation are:
\eqn\accoupl{
S_{int} = \int d^4x \sqrt{g_4(x,z_b)} \Bigl(
g^2_5 ~\Phi^2_b(x) ~\Phi^2_B(x,z_b) +
g_5~ \Phi_b(x)~ \lambda_b(x)~ \lambda_B(x,z_b)
\Bigr)\ ,}
where $\Phi_b, \lambda_b$ are brane scalars and fermions
respectively; $\Phi_B, \lambda_B$ are bulk scalars and fermions
at the brane; and $z_b$ is the location of the left IR brane
in the coordinates of \metricL.
In deriving the effective $4D$ interactions,
we normalize the fields so that they have
canonical kinetic terms in the $4D$ metric:
\eqn\rnorm{\phi_{4D~i} = a^{d_i} \Phi_{i}\ .}
Here $a = L/(|z|+L)$
is the scale factor in conformally flat coordinates;
$d_i$ is the mass dimension of $\phi_i$;
and $\Phi$ can be a bulk or brane field.
Next we perform the mode expansion for the bulk fields,
\eqn\modeexp{\eqalign{
&\phi_B(x,z) = \sum_{m}  \phi_{Bm}(x) \Psi_m(z) \cr
&\lambda_B(x,z) = \sum_{m} \lambda_{Bm}(x) \Psi_m(z)}\ .}
$\Psi_m$ are solutions of the Schr\" odinger equation
\eqn\schroeqs{\Psi''_m + (m^2 - V(z)) \Psi_m = 0\ ,}
with the potential $V$ determined by the spin and bulk couplings
of $\phi_B$.  As a function of $z$,
the bulk wavefunctions behave as
simple waves in a box.  They
depend explicitly on the locations
of the IR branes in the conformally flat coordinates;
and only implicitly on the AdS radius.

Next we expand the
action \accoupl\ in terms of the $4D$ modes.
In addition to the renormalizations above,
we must also take into account the
overlap between the bulk modes and the brane, $\Psi_B(y_b)$
in \accoupl. This can be calculated by computing the
normalization for the bulk modes
which reside between the left IR and the UV brane.
Wavefunctions for light, left-localized states are
linear combinations of Bessel functions:
$$\Psi_m = {\cal N} \sqrt{m(z+L)}
\{J_\alpha[m(z+L)] + (mL)^\alpha N_\alpha[m(z+L)]\}\ .$$
Far from the UV brane they asymptote
to trigonometric functions.  The
normalization condition $\int dz |\Psi_m|^2 = 1$
implies that
${\cal N} = 1/\sqrt{l}$, where $l$ is the conformal distance
between the IR and UV branes.
If we take the SM particles to be localized on the left IR brane,
the bulk wavefunction is:
$\Psi_m(y_b) \sim 1/\sqrt{l}$.
The $4D$ couplings can be determined by the equation:
\eqn\fdcoup{g^2_4 ~\phi_b^2 ~\phi_{Bm}^2 = a^4(z_b)
~g_5^2~ \Phi_b^2 ~\Phi^2_{Bm}(z_b)\ .}
Since the warp factor on the left IR brane
is $a(0)\simeq L/l$,
\eqn\scales{\eqalign{
&\Phi_b = {1 \over a(0)} ~\phi_b = {l\over L} \phi_b \cr
&\Phi_{Bm} = {\Psi_m(0)\over a^{3/2}(0) } ~\phi_{Bm}
\simeq  {l\over L^{3/2}} ~\phi_{Bm}\ .}}
If $g_5 \simeq 1/\sqrt{M_5}$, then:
\eqn\fdc{g_4 \simeq {1 \over \sqrt{M_5 L}\ .} }
The effective $4D$ interactions are:
\eqn\lagcoupl{
{\cal L}_{int} = \sum_m \Bigl( g^2_4
~\phi^2_b(x) ~\phi^2_{Bm}(x) + g_4~ \phi_b(x)
~ \lambda_b(x)~ \lambda_{Bm}(x)\Bigr) }
in canonically normalized field variables.

The annihilation cross section of the bulkies is:
\eqn\ancs{
\sigma \simeq N ~\Bigl({g_4^2 \over 4\pi}\Bigr)^2 ~{1\over m^2} \simeq
{N \over 16\pi^2  (M_5L)^2 m^2} \, .}
Using \density, and the virial velocity of DM particles $v \simeq 10^{-3} c$,
the relic density of the bulkies is therefore
\eqn\denbulk{{\rho \over \rho_c} \simeq 16\pi^2 \times 10^{-2} (M_5
L)^2 {m^2 \over TeV^2} \, .}
From this we see that there is a range of parameters for
which left bulkies near the electroweak scale can provide
the required dark matter density. For example, if $L M_5\sim 10$ and $m
\sim 100 GeV$, then the left bulkies close the universe.

Now, using Eq. \rel\ and assuming for simplicity that the
Calabi-Yau volume contribution to $M^2_4$ is comparable to
the throats, we find that the value of the AdS radius that
leads to a lifetime of the order of the age of the universe
is $L^{-1} \sim  10^{14} GeV$.  We should then take
$M_5 \sim 10/L \sim  10^{15} GeV$.
Finally, assuming
down the AdS throat that the extra five dimensions
are an Einstein manifold of volume $L^5$,
$M_{10} \sim (M_5^3/L^5)^{1/8} \sim 2.3 \times 10^{14} GeV$.
These rough numbers are within a few
orders of magnitude of those expected from
standard heterotic string model building, and are
close to the scale of coupling unification in the MSSM.
This rough coincidence suggests that
cold dark matter may be decaying into hot CFT dark
matter in our epoch; the hot DM then escapes
our galaxy. Such a conversion may have observational
implications: it can lead to a softening of the dark matter
density profile within our galactic halo by spreading it
into extragalactic space.  (This is the result, for example,
of simulations of decaying dark matter performed by Cen \cen).

What about the more economical possibility that the CDM
particle is the graviton? In contrast to the bulky, the
graviton is not protected by B-parity against decay into SM
particles. These decays would be much faster than the slow
tunneling into right-gravitons. One way to avoid this is to
place the SM sector on the Planck brane in the middle.
To see this we consider
the $4D$ effective action for massive gravitons coupled
to Standard Model fields, in the
linearized approximation. As in Sec. 2, write the
graviton fluctuations as:
$g_{\mu\nu} \rightarrow g_{\mu\nu} + [L/(|z|+L)]^{3/2}
\Psi_{m}
\gamma_{\mu\nu}$, where $\gamma_{\mu\nu}$
is the transverse, traceless $4D$ tensor and $\Psi_m$ is determined by
\schroeqs.
The action for $\gamma$ is:
\eqn\effact{S_{effective} = \sum_m \Bigl\{
\int d^4x \Bigl(\gamma_{\mu\nu}
K^{\mu\nu ~ \alpha\beta}(\partial, m)
\gamma_{\alpha\beta}\Bigr)
- \int d^4x {\Psi_m(0) \over M^{3/2}_5} \gamma_{\mu\nu} T^{\mu\nu}_{SM}
\Bigr\}}
where $K^{\mu\nu ~ \alpha\beta}(\partial, m)$ is the inverse
propagator for massive $4D$ gravitons, $T^{\mu\nu}$ is the
stress-energy tensor for SM fields and we have again
normalized the fields to have canonical kinetic terms on
the UV brane.

The KK graviton wavefunctions are described by Bessel functions
as explained above, with index $\alpha=2$. Far from the UV brane
they are approximated by ordinary trigonometric functions,
and hence as before are properly
normalized by $\sim 1/\sqrt{l}$.
On the other hand, near the UV brane these wave
functions are dominated by the
Neumann functions, of order
\eqn\normpsi{\Psi(0) \sim \sqrt{m(z + L)} m^2L^2
N_2[m(z+L)]/\sqrt{l} \sim \sqrt{mL/l} \ .}
This provides additional suppressions for the couplings. Because
$a(0)=1$ on the UV brane, the
coupling of the low-lying left-localized massive $5D$ gravitons to
matter on the UV brane
is $g \sim \sqrt{(mL/l)}/M^{3/2}_5 \gamma_{\mu\nu} T^{\mu\nu}$. Hence
the decay rate into SM particles is given by \eqn\brratfr{
\Gamma_{CDM \rightarrow SM} \sim N g^2 m^3 \sim
{m^4 L^4 \over l (M_5 L)^3} N = {\Gamma_{CDM \rightarrow
CFT} \over (M_5 L)^3} N \, .}
Since the rank of the gauge group in the dual CFT is
$\CN^2 = (M_5 L)^3$, the branching
ratio is:
\eqn\brrat{
{\Gamma_{CDM \rightarrow SM} \over \Gamma_{CDM \rightarrow CFT}}
= {N \over \CN^2}\ .}
Using the numbers above, $M_5 L \sim 10$ implies $\CN\sim 1000$,
and since (MS)SM has on the order of a $100$ flavors, we find
${\Gamma_{CDM \rightarrow SM}}/{\Gamma_{CDM \rightarrow CFT}} \sim
0.1$.
Hence when SM is localized on the UV brane,
the main decay mode of CDM is into CFT hot dark matter,
simply because of the universality of their couplings via the
stress-energy tensors and the large
number of degrees of freedom in the CFT.

However, since there are no particles on the left-brane, the
left-graviton has a small annihilation cross-section into the now
distant SM particles or the right-gravitons. To determine this, we
need the graviton-matter couplings which are quadratic in the
graviton field. It is clear that in terms of the graviton field
$\gamma_{\mu\nu}$ defined previously, these are
\eqn\actgqud{ S_{int}=\sum_m \Bigl\{\int d^4 x \sqrt{g_4}
{\Psi^2_m(0) \over M^{3}_5}\Bigl(
{\cal O}(1) \gamma_{\lambda\mu}\gamma^{\lambda}{}_{\nu} T^{\mu\nu}_{SM} +
{\cal O}(1) \gamma_{\mu\nu}\gamma^{\mu\nu} T_{SM} \Bigr) \Bigr\}}
where $T_{\mu\nu}$ and $T$ are the brane-localized SM
stress-energy tensor and its trace. Since $\Psi_m(0) \sim
\sqrt{mL/l}$, the quadratic graviton coupling is
\eqn\qgc{g \sim {mL \over l M^3_5}}
and because it has dimension $-2$, the annihilation cross section for the
process $\gamma \gamma \rightarrow SM$ is
\eqn\ancrsec{\sigma \sim g^2 m^2 \sim {m^4 \over l^2 M^6_5} L^2}
For the lightest KK gravitons, $m \sim 1/l$ and hence $\sigma \sim
(m/M_5)^6 L^2$. This cross section is so small that
when the gravitons are relativistic their annihilation
ceases, 
so that their ``freezeout''
abundance is comparable to that of photons. This means that
they are not cold but hot dark matter which is disfavored
by the observed structure formation at small scales. After
freezeout the gravitons are decoupled from the rest of the
particles whereas the photons continue to get ``heated-up''
at every threshold and, as a result, the photon abundance
today would be one to two orders of magnitude larger than that of
gravitons. Therefore in order that left-gravitons
close the universe today, they must weigh about $10 eV$.
So the conformal distance between the left IR brane and the UV brane
is $l \sim (10 eV)^{-1}$. Furthermore, in order
that its lifetime be the Hubble time the AdS radius must be
$L \sim GeV^{-1}$, and so $M_5 \sim 10^{13} GeV$.

An alternative possibility would be to postulate a time dependence in the
distance of the left brane from the UV brane which allows
for the annihilation cross section of the gravitons to be
large at some early epoch. This, or any other mechanism
enhancing the early graviton annihilation cross section
could allow for heavy CDM gravitons. However, this amounts
in practice to giving up calculability and simply
postulating the existence of a mechanism giving rise to the
correct CDM abundance for $\sim TeV$ gravitons.

Cosmological models with standard cold dark matter consisting of
weakly interacting massive particles (WIMPS) successfully account
for the large scale structure of the universe at extragalactic
scales. However, there have been suggestions of a number of problems at
subgalactic scales, associated with an apparent excess of small
scale structure. Simulations predict cuspy halo density
profile at subgalactic scales which give rise to galaxy and
cluster cores that appear to be overly dense. They also predict a
large number of subhalos which are disfavored by the scarcity of
satellite galaxies in clusters. There have been several proposals
to account for these discrepancies including collisional dark
matter \spesta, fluid DM \peebles, decaying DM \cen\ and so on.
The common theme is the presence of a mechanism
that spreads the localized CDM
energy into a larger region of space.

Our proposal is similar to these ideas and its implications
for structure formation are close to the scenario of
\cen\ where the CDM particle decays to light, invisible particles --
neutrinos, for example --
which escape our galaxy. There it was
shown by detailed computer simulations that if a
significant fraction (e.g., one-half) of the CDM decays into
neutrinos, then at least some of the problems with excess
small-scale structure are avoided. The only difference in
our case is that the decay product is CFT-matter instead of
neutrinos. A potential difficulty of this scenario is that
it predicts that younger galaxies should be heavier by an
amount which may be too large \foot{ We thank Paul
Steinhardt for pointing this out.}.

A variation of this scenario is to add a right IR brane. If
the characteristic length scale of this brane is larger
than a galactic halo $\sim Mpc$ the previous discussion is
not significantly altered: we are again dealing with an
approximate CFT. If the scale of the right brane is less
than a galactic halo then the IR dynamics of confinement
sets in before the right localized gravitons escape our
halo and may prevent them from leaving it. The detailed
computation of sub-galactic structure formation is expected
to be involved in this case. After all, even for the simple
dissipationless cold dark matter structure computation
gave surprising results, such as the excess subgalactic
structure we are attempting to undo. Nevertheless, it is
clear that the possibility of confining the right-gluons
within the galaxy will eliminate the problem of
heavier young galaxies mentioned earlier. Still, depending
on the scale of confinement, the decaying CDM model
suggests that faraway galaxies should have denser and more
compact cores because they are younger.

\newsec{Spectrum and Tunneling-Mediated Supersymmetry Breaking}

Low-energy SUSY is one of the most attractive scenarios for
physics beyond the Standard Model, because it stabilizes scalar
masses at the SUSY breaking scale \dg.
We must still explain
the origin of the low SUSY breaking
scale and the 125 physical parameters in the
MSSM \sutter.
The standard lore is to postulate that supersymmetry is broken in a hidden
sector, and that the breaking communicated to the visible sector
via some messenger fields.
There is a variety of SUSY breaking mediation mechanisms
such as gravity \refs{\suGUT,\nil,\gauco}, gauge \refs{\hs,\dine},
anomaly \rszero, gaugino \mart\ mediation,
in hidden sector scenarios.

In brane world models,
supersymmetry can be broken on a brane located
``elsewhere'' in the extra dimensions \hw\mirpe. The warped
compactifications provide a new
calculable regime for this kind of idea. Specifically
tunneling
effects between brane throats provide a new mechanism for
generating a small SUSY breaking scale.
Here we consider the following scenario.
We take both IR branes to be close to the UV
brane. SUSY is broken on the left IR brane via soft mass-like terms for bulk
modes. This induces
SUSY breaking mass splittings on the right IR brane, where the SSM
resides. A hierarchy is generated because SUSY breaking at a high
scale on the left IR brane induces small SUSY breaking mass
splittings in the SSM, due to tunnelling supression. We choose the
distance between the right IR brane and the UV brane, as well as
the bulk parameters $M_5$ and $L$, to be near the GUT scale
$M_{GUT} \sim 10^{16} GeV$. This ensures that the cutoff on the
right SSM brane is $M_{GUT}$; consequently supersymmetric gauge
coupling unification \refs{\dg,\drw}\ can be preserved.
We find that there is a regime where the $5D$ tunneling effects
could be the dominant channel for transmitting SUSY breaking
to the visible sector.

There may be different sources of SUSY breaking in the
distant throat.
The simplest way to break supersymmetry is by considering
a bulk hypermultiplet with a massless
scalar $\phi$, and breaking supersymmetry explicitly with a
mass-like term for $\phi$. We will consider this example first,
and later will consider a more realistic but involved
case when SUSY is broken by Majorana-like mass terms for
bulk fermions, which is necessary to generate gaugino masses.

In order to separate the effects of hierarchy generation
by warp factor scaling as in \RSI\ from the
effects of the tunneling, we consider two cases:

\noindent
{\it 1) Direct SUSY breaking}: We live on
the left IR brane at $z=-l_1$,
and SUSY is broken in the left throat. The effects
of the right throat on SUSY breaking are negligible.
This case provides a
contrast for the next case, which is our main focus.

\noindent
2) {\it Tunneling mediation}: In the
two-throat geometry of Figure 2, supersymmetry is broken in
the left throat, and our world is right IR brane at $z = l_2$.

\noindent We then compute the mass splittings these effects
induce on particles
localized on the relevant branes: the left IR brane for case 1),
and the right IR brane for case 2).

We discuss
two (qualitatively different) types of supersymmetry breaking:

\noindent
{\it Type A)}
SUSY is broken
everywhere in the left throat, for example by taking a IIB string
theory background
of the form $AdS_5 \times X$, where $X$ is an Einstein manifold or
deformation thereof which
breaks supersymmetry (simple
examples, from orbifolds or RG flows starting
from $AdS_5 \times S^5$, appear in
\adsbreak). We model this with
a step function mass
squared for bulk particles $m^2_0 + \bar q^2$ for $z < 0$
and $m^2_0$ for $z > 0$, where only the
$\bar q^2$ piece breaks supersymmetry (\ie\ it is
zero for the superpartner.)

\noindent
{\it Type B)}
Supersymmetry-breaking is localized
on the left IR brane.
In the AdS/CFT correspondence, this corresponds to a case
where the gauge theory dual to closed
strings in the left throat
dynamically breaks supersymmetry,
and so is supersymmetric down to a low scale.
We model this with a
localized, $\delta$-function mass on the left brane.

In \S4.1\ we consider the spectra of bulk modes in the presence of the SUSY
breaking terms of types A) and B).
We also discuss the decomposition into modes localized in
the left and right throat, and delineate the approximations
made in the process.
In \S4.2\ we discuss type A) SUSY breaking
in detail and determine
the couplings of the left and right localized bulk modes to the
right IR brane.
In \S4.3, we redo this for type B) SUSY breaking.
In \S4.4 we calculate the splittings of
superpartners on the right IR brane.
We also find the splittings on the left IR brane in the
various cases, to compare type 1) and type 2) SUSY breaking.

These calculations illustrate how SM
matter fields obtain their SUSY breaking masses.
To understand gaugino masses, we must break R-symmetry,
and we give an example in \S4.5.
We discuss the parameter values of the 5d model
which would lead to phenomenologically interesting and
viable SUSY breaking in \S5.

\subsec{Spectra of Bulk Modes with Broken SUSY}

We study a scalar field $\phi$ with supersymmetric
bulk mass
$m_0$ propagating in the background metric \metricL.
The IR branes function as orbifold planes
for the purposes of defining boundary conditions for
bulk fields, as in \refs{\hw,\RSI}.
We will assume that$l_1 > l_2$ both for model building purposes and
in order to use
nondegenerate perturbation theory
for determining the spectra of bulk modes. If we had
taken $l_1 =l_2$ instead, the low-lying modes
would have had nearly degenerate states, even or odd under
$z \rightarrow - z$. However $l_1 > l_2$ automatically
selects the left- and the right-localized states as the natural basis
of bulk modes.
We can then immediately determine
the mass splitting for all bulk states, avoiding
the complications arising from the near-degeneracy
of symmetric and antisymmetric modes when $l_1= l_2$.

With SUSY breaking mass terms $M^2_0 =m^2_0 + \bar q^2 \theta(-(z+l_1))$
and $2q$ (where $\bar q$ parameterizes the type A) breaking and $2q$
the type B) breaking),
the scalar equation is
\eqn\scaleq{
{1\over \sqrt{-G}} \partial_{a}(\sqrt{-G} G^{ab}
\partial_{b})\phi~=~\Bigl(M_0^2 + 2 q \sqrt{G_4 \over G}
\delta(z+l_1) \Bigr) \phi\ ,
}
where $G$ is the $5D$ metric, $G_4$ is the induced $4D$ metric and
the indices $a,b$ run over all 5 dimensions.   Making
the KK ansatz appropriate for modes of fixed 4d mass $m$
(which satisfy the $4D$ Klein-Gordon
equation $\partial^2_4 \phi = m^2 \phi$), and defining
\eqn\newfield{\psi = ({L \over |z|+L})^{3/2} \phi\ ,}
one finds that $\psi$ satisfies Bessel's equation
\eqn\bessel{\psi'' + \left( m^2 - {15+ 4M_0^2 L^2 \over 4(|z|+L)^2}
- 2 q {L \over l_1+L} \delta(z+l_1) \right) \psi ~=~0\ ,}
where we take the supersymmetric brane-localized terms to
vanish, since they do not alter the conclusions qualitatively.
Using the new variable
\eqn\newvar{w = m(|z|+ L)}
we can rewrite the mode equation \bessel\ in the bulk as
\eqn\modeeq{{d^2 \psi \over dw^2} + \left(1 -
{15 + 4M^2_0 L^2 \over 4 w^2} \right) \psi = 0}
The solutions are
\eqn\psileft{\psi(w) = \psi_L(w) = \sqrt{w}
\left( a_L J_{\nu_L}(w) + b_L J_{-\nu_L}(w)\right),
~~~~~ z<0}
\eqn\psiright{\psi(w) = \psi_R(w) = \sqrt{w}
\left( a_R J_{\nu_R} (w) + b_R J_{-\nu_R} (w) \right),
~~~~~ z>0}
where $J_\alpha$
are the Bessel functions of index $\alpha$, and where
\eqn\nudef{\nu ~=~\sqrt{4 + M_0^2 L^2 }}
Note that the indices of the Bessel functions differ in the two
throats, due to the step function mass.
We have taken $\nu_{L,R}$ to be noninteger
numbers to simplify the calculations; otherwise
$J_{-\alpha}$ should be replaced by Neumann functions
$N_{\alpha}$.  This assumption will not affect our final result.

The wavefunctions satisfy the boundary conditions
\eqn\bcs{\eqalign{
&\psi'(-l_1) = q {L \over l_1+L} \psi(-l_1),
~~~~~~~~~~~~~~~~~~~~~~~~ left~ IR ~ brane \cr
&\psi(0_+) = \psi(0_-), ~~~~ \psi'(0_+) = \psi'(0_-),
~~~~~~~~~~ UV ~ brane \cr
&\psi'(l_2) = 0, ~~~~~~~~~~~~~~~~~~~~~~~~~~~~~
~~~~~~~~~~~~~~~right ~IR ~ brane\cr } }
The first equation comes from the $\delta$-function SUSY breaking
mass term on the left IR brane. In the absence of this term,
the wavefunction would satisfy Neumann boundary conditions
on the left IR brane,
as it does on the supersymmetric IR brane on the right.
In general, the boundary conditions corresponding to the
supersymmetric limit may be different, but that does
not change our qualitative conclusions.

The solutions depend on five integration parameters $a_{L,R},b_{L,R}, m$.
Removing the overall scale of the
wavefunction, and imposing the four boundary conditions at $z=0$
and $z_1= -l_1$, $z_2 = l_2$ \bcs\ , we find a discrete
tower of solutions for
$m$. We now discuss separately the
cases when supersymmetry is broken
in the whole left throat and
when it is broken only on the left IR brane.

\subsec{SUSY Breaking in the Left Throat}

Let us first set $\bar{q} \neq 0$, $q = 0$.
The superpartner has mass $m_0$, so SUSY is unbroken
in the right throat.
We compute the mass splittings in the
4d Kaluza-Klein tower and the
effective 4d couplings to the fields on the IR branes.

For the case $\bar q << m_0$,
\eqn\nus{\eqalign{
&\nu_R = \nu = \sqrt{4 + m^2_0L^2} \cr
&\nu_L = \sqrt{4 + M^2_0 L^2} \sim \nu + {\bar q^2 L^2 \over 2\nu} \cr}\ .}
This approximation is very good because of the quadratic dependence
of $\nu_L$ on $\bar q$.

Substituting the solutions \psileft, \psiright\ into the eqs. \bcs\
gives
\eqn\stepsys{\eqalign{
&b_R = \alpha a_R + \beta a_L ~~~~~~~~~~
b_L = \bar \beta a_R + \bar \alpha a_L \cr
&b_R = \gamma a_R ~~~~~~~~~~~~~~~~~~~~
b_L = \bar \gamma a_L }}
where the coefficients are
\eqn\stepcoef{\eqalign{
&\alpha = {{J_{\nu_R-1}(mL) - (\nu_R -1/2) J_{\nu_R}(mL)/mL \over
J_{-\nu_L+1}(mL) + (\nu_L -1/2) J_{-\nu_L}(mL)/mL } - {J_{\nu_R}(mL)
\over J_{-\nu_L}(mL)}
\over
{J_{-\nu_R+1}(mL) + (\nu_R -1/2) J_{-\nu_R}(mL)/mL \over
J_{-\nu_L+1}(mL) + (\nu_L -1/2) J_{-\nu_L}(mL)/mL } + {J_{-\nu_R}(mL)
\over J_{-\nu_L}(mL)}} \cr
&\beta = {{J_{\nu_L-1}(mL) - (\nu_L -1/2) J_{\nu_L}(mL)/mL \over
J_{-\nu_L+1}(mL) + (\nu_L -1/2) J_{-\nu_L}(mL)/mL } + {J_{\nu_L}(mL)
\over J_{-\nu_L}(mL)} \over
{J_{-\nu_R+1}(mL) + (\nu_R -1/2) J_{-\nu_R}(mL)/mL \over
J_{-\nu_L+1}(mL) + (\nu_L -1/2) J_{-\nu_L}(mL)/mL } + {J_{-\nu_R}(mL)
\over J_{-\nu_L}(mL)}} \cr
&\bar \alpha = {{J_{\nu_L-1}(mL) - (\nu_L -1/2) J_{\nu_L}(mL)/mL \over
J_{-\nu_R+1}(mL) + (\nu_R -1/2) J_{-\nu_R}(mL)/mL } - {J_{\nu_L}(mL)
\over J_{-\nu_R}(mL)}
\over
{J_{-\nu_L+1}(mL) + (\nu_L -1/2) J_{-\nu_L}(mL)/mL \over
J_{-\nu_R+1}(mL) + (\nu_R -1/2) J_{-\nu_R}(mL)/mL } + {J_{-\nu_L}(mL)
\over J_{-\nu_R}(mL)}} \cr
&\bar \beta = {{J_{\nu_R-1}(mL) - (\nu_R -1/2) J_{\nu_R}(mL)/mL \over
J_{-\nu_R+1}(mL) + (\nu_R -1/2) J_{-\nu_R}(mL)/mL } + {J_{\nu_R}(mL)
\over J_{-\nu_R}(mL)} \over
{J_{-\nu_L+1}(mL) + (\nu_L -1/2) J_{-\nu_L}(mL)/mL \over
J_{-\nu_R+1}(mL) + (\nu_R -1/2) J_{-\nu_R}(mL)/mL } + {J_{-\nu_L}(mL)
\over J_{-\nu_R}(mL)}} \cr
&\gamma = {J_{\nu_R-1}(m(l_2+L)) - (\nu_R -1/2) J_{\nu_R}(m(l_2+L))/m(l_2+L) \over
J_{-\nu_R+1}(m(l_2+L)) + (\nu_R -1/2) J_{-\nu_R}(m(l_2+L))/m(l_2+L) } \cr
&\bar \gamma = {J_{\nu_L-1}(m(l_1+L)) - (\nu_L -1/2)
J_{\nu_L}(m(l_1+L))/m(l_1+L) \over
J_{-\nu_L+1}(m(l_1+L)) + (\nu_L-1/2) J_{-\nu_L}(m(l_1+L))/m(l_1+L) }  \ .}}
Eqs. \stepsys\ are four homogenous equations in four unknowns;
for a solution to exist the coefficients \stepcoef\ must
satisfy
\eqn\eigen{(\gamma - \alpha)(\bar \gamma - \bar \alpha) - \beta \bar \beta
 = 0\ .}
Combined with Eq. \stepcoef, the roots determine the
mass eigenvalues in the spectrum.

The light bulk states which contribute to SUSY breaking
satisfy $mL << 1$, $ml_k >> 1, k=1,2$.
Therefore, at $z = 0$ we can approximate Bessel functions as
$J_\nu(x) \simeq (x/2)^\nu/\Gamma(\nu+1)$.
Eq. \stepcoef\ becomes:
\eqn\apstepcoef{\eqalign{
&\alpha = {\cal O}(1) (mL)^{2\nu_R} ~~~~~~~~~~
\beta = {\cal O}(1) (mL)^{\nu_L+\nu_R} \cr
&\bar \alpha = {\cal O}(1) (mL)^{2\nu_L} ~~~~~~~~~~
\bar \beta = {\cal O}(1) (mL)^{\nu_L + \nu_R} \ .}}
where ${\cal O}(1)$ denotes unimportant dimensionless parameters
of order unity. The eigenvalue equation \eigen\ simplifies to
\eqn\apeigen{ \gamma \bar \gamma - {\cal O}(1) (mL)^{2\nu_L} \gamma
- {\cal O}(1) (mL)^{2\nu_R} \bar \gamma - {\cal O}(1)
(mL)^{2\nu_L + 2\nu_R} =0}
The information about SUSY breaking is fully encoded in the parameter
$\nu_L$, as is evident from \nus.

The spectrum for low 4d masses contains states roughly
localized on each side of the UV brane.
Eqs. \stepsys,\apstepcoef\
imply that the states  localized on the left are characterized by
$\bar \gamma \simeq 0, \gamma \ne 0$.
Therefore in \apeigen\ we can ignore the last two terms on the LHS;
cancelling $\gamma$ yields $\bar \gamma = {\cal O}(1)(mL)^{2\nu_L}$.
But since $mL << 1$, SUSY breaking from $\nu_L$ on the RHS is miniscule
compared to SUSY breaking arising from the explicit presence of $\nu_L$
in the formula for $\bar \gamma$ in
\stepcoef\ . Thus to
leading order the SUSY-breaking spectrum of left-localized modes
satisfies
\eqn\lefteigen{ \bar \gamma = 0\ .}
Using $ml >> 1$ near the IR branes we can use
the large argument asymptotics of the Bessel functions. Since
${\cal O} (L/l_1) << 1$, \lefteigen\ leads to
\eqn\lefteigsin{ \sin[ml_1 - (\nu_L + 1/2)\pi/2] = -
{\nu_L - 1/2 \over ml} \cos[ml_1 - (\nu_L + 1/2)\pi/2]\ . }
The supersymmetric masses take the form
\eqn\susmass{ m = {\cal O}(1) {n - \beta \over l_1}\ ,}
where $n$ is an integer and $\beta$ is a fraction which depends only
on the supersymmetric parameters and is roughly the same for all states.
This formula for the supersymmetric mass
will remain true for all the bulk states in what follows.
Eqs. \lefteigsin,\nus\ give the mass
splitting between the scalars and their superpartners, to leading
order in $\bar q$:
\eqn\splitleft{\delta m_L = {\cal O}(1) {\bar q^2 L^2 \over l_1}}

For left-localized modes,
$b_L \sim (mL)^{2\nu} a_L$,
$b_R \sim (mL)^{2\nu} a_L$ and $a_R \sim (mL)^{2\nu} a_L$,
so the dominant contribution to the wavefunction comes from the mode
$\sim a_L$ localized in the region $z < 0$. For $ml >> 1$
the Bessel functions can be approximated by
trigonometric functions for most of their support in $z$, and so
the normalization condition
$\int dz |\psi|^2 = 1$ requires that
$l_1 a_L^2 + l_2 a_R^2 = {\cal O}(1)$. This yields
$a_L \sim {\cal O}(1) (l_1 + l_2 (mL)^{4\nu})^{-1/2}$.
Since $l_1 > l_2$ and $mL <1$, we find $a_L \sim {\cal O}(1) l^{-1/2}_1$
to be an excellent approximation for all left-localized modes
which contribute to SUSY breaking.

The overlap of the left-localized modes with the
left IR brane is
$\sim a_L \sim 1/\sqrt{l_1}$, while the overlap with
the right IR brane is
$\sim a_R \sim (mL)^{\nu_L+\nu_R}/\sqrt{l_1} \sim (mL)^{2\nu}/\sqrt{l_1}$.
We renormalize the fields as in sec. 3, to make
the kinetic terms canonical.
Again, the proper rescalings leading to
correct couplings of canonical $4D$ fields
on the $j^{\rm th}$-brane are:
for bulk scalars $\Phi_B  = \psi/a(l_j)^{3/2}\phi_B \sim \psi
(l_j/L)^{3/2} \phi_B$ and for brane fields
$\Phi_b = \phi_{4D}/a(l_j) \sim \phi_{4D} l_j/L$.
A generic interaction term on the brane is:
$(g^{I}{}_{J~5D})^2
(\Phi^{J}{}_b)^2 (\Phi^{I}{}_B)^2$,
where $I,J$ are indices denoting left ($L$) and right ($R$).
The effective $4D$ couplings
of the  left-localized bulk states $g^L{}_{I~4D} \sim g^L{}_I$
\eqn\couplingleft{\eqalign{
& g^L{}_{L~4D} = {g^L{}_L \over \sqrt{L}} ~~~~~~~~~~~~~~~~~~~~~~~~~~~
{\rm to~ left ~ IR ~ brane ~ states} \cr
& g^L{}_{R~4D} = {g^L{}_R (mL)^{2\nu}
\over \sqrt{L}} \sqrt{l_2 \over l_1} ~~~~~~~~~~~
{\rm to~ right ~ IR ~ brane ~ states} \cr
}}
The couplings to the right IR brane states are
very small, due to the high power $(mL)^{2\nu}$, $mL << 1$.
This suppression is a manifestation of tunneling.

The right-localized states are characterized by
$\gamma \simeq 0, \bar \gamma \ne 0$.
Since we are interested in the mass splitting $\delta m$
between the
scalars and their superpartners to leading order in $\bar q$, we
need to solve
$\gamma = {\cal O}(1)(mL)^{2\nu_L+2\nu_R}/\bar \gamma = {\cal O}(1)
(mL)^{4\nu}
\bar q^2 L^2$. Substituting the asymptotic formula for Bessel functions into
$\gamma$ in \stepcoef\ we obtain
\eqn\righteigen{
\sin[ml_2 - (\nu + 1/2)\pi/2] +
{\nu - 1/2 \over ml_2} \cos[ml_2 - (\nu + 1/2)\pi/2] = {\cal O}(1) (mL)^{2\nu} +
{\cal O}(1) (mL)^{4\nu}
\bar q^2 L^2}
leading to the supersymmetric masses similar to \susmass,
\eqn\susmassr{ m = {\cal O}(1) {n - \beta \over l_2}\ ,}
and
SUSY-breaking mass splitting
\eqn\splitright{\delta m_R = {\cal O}(1) ~(mL)^{4\nu}
~{\bar q^2 L^2 \over l_2}}
The mass splitting for the right-localized states is much
smaller, by the factor $(mL)^{4\nu}$,
than for the left-localized states \splitleft.
This is an effect
of the tunneling suppression
of the wavefunction in the left throat,
as seen in the relations $b_L \sim (mL)^{2\nu} a_R$,
$b_R \sim (mL)^{2\nu} a_R$ and $a_L \sim (mL)^{2\nu} a_R$.
From the form of the Bessel functions
for large arguments, the correct normalization of the wavefunctions
requires $l_1 a^2_L + l_2 a^2_R = {\cal O}(1)$.
This and $a_L \sim a_R (mL)^{2\nu}$ yields
$a_R \sim {\cal O}(1) (l_2 + (mL)^{4\nu} l_1)^{-1/2}$.
Note that since $l_2 < l_1$ the two terms in the normalization formula compete.
When $l_1 > l_2 (mL)^{-4\nu}$ the left side contributions dominate,
and hence they must not be neglected in general.
Their presence indicates that when the
space on the right of the barrier is tiny compared
to that on the left side, it does not
localize bulk modes efficiently. Those  modes leak to the left.

The overlap of the  right-localized modes with the
left IR brane is $\sim a_L \sim (mL)^{2\nu} /\sqrt{l_2 + (mL)^{4\nu} l_1}$,
and with
the right IR brane it is $\sim a_R \sim 1/\sqrt{l_2 + (mL)^{4\nu} l_1}$.
As before, the effective $4D$ theory couplings
of the  right-localized bulk modes $g^R{}_{i~4D}$ are given by
\eqn\couplingright{\eqalign{
& g^R{}_{L~4D} = {g^R{}_L (mL)^{2\nu} \over \sqrt{L}}
\sqrt{l_1 \over l_2 + (mL)^{4\nu} l_1} ~~~~~~~~
{\rm to~ left ~ IR ~ brane ~ states} \cr
& g^R{}_{R~4D} = {g^R{}_R \over \sqrt{L}}
\sqrt{l_2 \over l_2 + (mL)^{4\nu} l_1} ~~~~~~~~~~~~~~~~~~
{\rm to~ right ~ IR ~ brane ~ states} \cr
}}
The couplings to the left IR brane states are now
very suppressed due to tunneling as long as $l_1 < l_2 (mL)^{-4\nu}$,
which is the condition for strong localization of right modes.

What about the heavy states with $mL > 1$? For these states the
barrier is lower than their (bulk) energy and therefore they can
fly over it without significant suppression. Hence
they are not strongly localized on different sides of the UV brane,
but are only slightly asymmetric. This means that the
tunneling suppression factor $\sim (mL)^{2\nu}$ quickly converges to unity for
these states, once the additional corrections in the $mL$
expansion are summed up. Hence the masses and the splittings of the left- and right-side
modes become roughly the
same, and are controlled by the size of the conformal box
inhabited by the bulk modes, $l_1+l_2$:
\eqn\heavyms{\eqalign{
&m = {\cal O}(1) {n - \beta \over l_1+l_2}\cr
&\delta m = {\cal O}(1) {\bar q^2 L^2 \over l_1 + l_2}}}
while the couplings are
\eqn\heavycouplings{\eqalign{
&g^L{}_{L~4D} ={g^L{}_L \over \sqrt{L}} \sqrt{l_1 \over l_1 + l_2}
~~~~~~~~~~~~~~~~~~
g^L{}_{R~4D} = {g^L{}_R \over \sqrt{L}} \sqrt{l_2\over l_1 + l_2}
\cr
&g^R{}_{L~4D} ={g^R{}_L \over \sqrt{L}} \sqrt{l_1 \over l_1 + l_2}
~~~~~~~~~~~~~~~~~~
g^R{}_{R~4D} = {g^R{}_R \over \sqrt{L}} \sqrt{l_2\over l_1 +
l_2}}}
Here the numerator of the square root comes from the
conformal rescaling by the powers on the warp factor, and is
determined by the location of the brane, whereas the denominator
is the normalization factor of the bulk wave function. The
calculation is straightforward along the lines for the light
states. This completes the construction
of the bulk spectra when SUSY is broken in the
entire left throat.

It is already apparent that the left-localized modes
dominate direct SUSY breaking (type 1): these modes
have large mass splittings and large couplings to
the left IR brane, while the right-localized modes
have small mass splittings and small coulings to the
left IR brane.

On the other hand, both the left- and right-localized
bulk modes contribute to SUSY breaking on the right IR brane.
The left-localized modes have small couplings
$g^L{}_{R~4D}$ to the right IR brane states,
and the right-localized
states are only weakly split.
The mass splitting transmitted to
the right IR brane will therefore be small
We will compute the
transmitted mass splittings explicitly and isolate the leading
effects. But first, we determine the spectrum of bulk states
if SUSY is broken only on the left IR brane.

\subsec{SUSY Breaking on the Left IR Brane}

Now we set $\bar q = 0$, $q \neq 0$.  In this case,
\eqn\brnus{ \nu_L = \nu_R = \nu = \sqrt{4+m^2_0 L^2}\ .}
The substitution of the solutions \psileft, \psiright\
into the boundary conditions \bcs\ with $\bar q = 0, q \ne 0$
relates the constants
$a_{L,R}, b_{L,R}$ in the same way as in \stepsys,
except that now the explicit formulas for the coeficients
in terms of Bessel functions are different.
Define
\eqn\hatq{ \hat q = q {L \over l_1+L} }
which is well approximated by $\hat q \simeq q {L\over l}$
for $l/L >> 1$.  We find
\eqn\brcoef{\eqalign{
&\alpha = \bar \alpha = {1 \over 2}\Bigl[{J_{\nu-1}(mL)
- (\nu -1/2) J_{\nu}(mL)/mL \over
J_{-\nu+1}(mL) + (\nu -1/2) J_{-\nu}(mL)/mL } - {J_{\nu}(mL)
\over J_{-\nu}(mL)} \Bigr] \cr
&\beta = \bar \beta = {1 \over 2} \Bigl[{J_{\nu-1}(mL)
- (\nu -1/2) J_{\nu}(mL)/mL \over
J_{-\nu+1}(mL) + (\nu -1/2) J_{-\nu}(mL)/mL } + {J_{\nu}(mL)
\over J_{-\nu}(mL)} \Bigr]\cr
&\gamma = {J_{\nu-1}(m(l_2+L)) - (\nu -1/2) J_{\nu}(m(l_2+L))/m(l_2+L) \over
J_{-\nu+1}(m(l_2+L)) + (\nu -1/2) J_{-\nu}(m(l_2+L))/m(l_2+L) } \cr
&\bar \gamma = {J_{\nu-1}(m(l_1+L)) - (\nu -1/2) J_{\nu}(m(l_1+L))/m(l_1+L) +
\hat q  J_{\nu}(m(l_1+L))/m \over
J_{-\nu+1}(m(l_1+L)) + (\nu-1/2) J_{-\nu}(m(l_1+L))/m(l_1+L) -
\hat q  J_{-\nu}(m(l_1+L))/m  }  \cr }}
The coefficients must still satisfy a consistency condition,
leading to the eigenvalue equation \eigen\ as before.

For light states $mL << 1$, $m l_k > 1$ we
again approximate the Bessel functions at the
IR and UV branes by their asymptotic
values for large and small arguments respectively. As before,
\eqn\brcoefnos{ \alpha = {\cal O}(1) (mL)^{2\nu} ~~~~~~~~~~~~~~~~~~~~~
\beta = {\cal O}(1) (mL)^{2\nu} }
and therefore the eigenvalue equation becomes
\eqn\eigenbr{  \gamma \bar \gamma - {\cal O}(1) (mL)^{2\nu} (
\gamma + \bar \gamma) - {\cal O}(1) (mL)^{4\nu} = 0 }
Now the SUSY breaking scale is completely contained within $\bar
\gamma$.

The bulk
spectrum again naturally splits into
left- and right-localized states. The left-localized
states again have $b_L \sim (mL)^{2\nu} a_L$, $b_R \sim (mL)^{2\nu}
a_L$, $a_R \sim (mL)^{2\nu} a_L$, and
$\bar \gamma \simeq (mL)^{2\nu} << \gamma \ne 0$.
Because we wish to detemine the mass
splitting $\delta m$ to leading order in $\hat q$, it is sufficient
to consider $\bar \gamma = 0$, or
\eqn\leftbreig{J_{\nu-1}(ml_1) = \Bigl(
 {\nu - 1/2 \over ml} - {\hat q \over m} \Bigr) J_{\nu}(ml_1) }
which at the IR brane can be approximated by:
\eqn\lefteigapp{\sin[ml_1 - (\nu + 1/2)\pi/2] =
\Bigl({\hat q \over m} - {\nu - 1/2 \over ml_1} \Bigr) \cos[ml_1 - (\nu +
1/2)\pi/2]}
The supersymmetric masses are the solutions of this equation
when $q = 0$, and are given by Eq. \susmass.
We will find below that $q$ is within an order of magnitude or two of
$1/L$.  Since $\hat q l_1 \simeq q L$, the ratio $\hat q/m < 1$
for some of the relevant states, and
$\hat q/m > 1$ for the rest. These cases must be treated
separately in Eq. \lefteigapp. For the heavy states,
the SUSY breaking terms are small perturbations,
so that $\delta m \sim \hat q/(ml_1)$. For the light states,
the SUSY breaking terms are a large perturbation,
leading to a change in the mass of order of ${\cal O}(1)/l_1$.
The results for the mass splittings are
\eqn\dmassl{\eqalign{
& \delta m_L = {\cal O}(1) {1 \over l_1 } ~~~~~~~~~~~~ {\rm when ~}
m < \hat q \cr
& \delta m_L = {\cal O}(1) {\hat q \over l_1m} ~~~~~~~~~~ {\rm when ~}
m > \hat q \cr }}
The effective four-dimensional
couplings are the same as in type A) breaking
as summarized in Eq. \couplingleft.

The masses, splittings, and couplings of the
right-localized states are computed in the same way.
For these states, $b_L \sim (mL)^{2\nu} a_R$, $b_R \sim (mL)^{2\nu}
a_R$, $a_L \sim (mL)^{2\nu} a_R$, $\gamma \simeq 0$, and
$\bar \gamma \ne 0$. Determining the mass splitting is
subtler than before. As $l_2 >> L$,
the eigenvalue equation is approximately:
\eqn\rbreigen{\eqalign{
&J_{\nu-1}(ml_2) -  {\nu -1/2 \over ml_2} J_{\nu}(ml_2) =
{\cal O}(1) (mL)^{2\nu} \cr
&~~~~~~~~~~~~ + {\cal O}(1) (mL)^{4\nu}
{J_{-\nu+1}(ml_1) + (\nu-1/2) J_{-\nu}(ml_1)/ml_1 -
\hat q  J_{-\nu}(ml_1)/m
\over
J_{\nu-1}(ml_1) - (\nu -1/2) J_{\nu}(ml_1)/ml_1 +
\hat q  J_{\nu}(ml_1)/m  }\ .}}
When $q =0$ the roots yield the masses of the light supersymmetric
modes, which are of the same form as \susmassr.
Solving perturbatively for $\delta m$ as a function of $\hat q$,
we find that for $\hat q >> m$
\eqn\rbreigenoneg{
J_{\nu-1}(ml_2) - {\nu -1/2 \over ml_2} J_{\nu}(ml_2) =
{\cal O}(1) (mL)^{2\nu} + {\cal O}(1) (mL)^{4\nu}\ ,}
where the first term on the RHS is a supersymmetric shift
of the mass $\sim (mL)^{2\nu}/l_2$ which merely
renormalizes $\beta$ in \susmass, while the second term arises
because of the SUSY breaking. Hence the mass splitting induced
by the SUSY breaking terms is
$\delta m \sim (mL)^{4\nu}/l_2$ when $\hat q/m > 1$,
again using the large argument asymptotics of
Bessel functions.
When $\hat q/m <1$, we find
\eqn\rbreigenonel{
J_{\nu-1}(ml_2) - {\nu -1/2 \over ml_2} J_{\nu}(ml_2) =
{\cal O}(1) (mL)^{2\nu} + {\cal O}(1) (mL)^{4\nu} +
{\cal O}(1) (mL)^{4\nu} {\hat q \over m}}
and so the SUSY breaking terms induce mass splitting
$\delta m \sim (mL)^{4\nu} \hat q/(ml)$. (Note that
the second term on the RHS of (\rbreigenonel) does
not arise from SUSY breaking, unlike the similar term
in Eq. (\rbreigenoneg).)  Summarizing, we find that
\eqn\dmassr{\eqalign{
& \delta m_R = {\cal O}(1) ~(mL)^{4\nu} ~{1 \over l_2}
~~~~~~~~~~~~ {\rm when ~} m < \hat q \cr
& \delta m_R = {\cal O}(1) ~(mL)^{4\nu} ~{\hat q \over l_2m}
~~~~~~~~~~ {\rm when ~} m > \hat q \cr }}

In this case the correct normalization of the wavefunctions
is $a_R = {\cal O}(1) (l_2+(mL)^{4\nu}l_1)^{-1/2}$.
Again for $l_1 > l_2 (mL)^{-4\nu}$ the left side contributions would
dominate, indicating that the modes with
such a high mass cease to be localized on the right.
The effective four-dimensional couplings
of the right-localized modes are identical
to that in type A) SUSY breaking, as seen in Eq. (\couplingright).

The couplings of the heavy states with $mL > 1$ behave qualitatively
the same as when
SUSY is broken in the whole left throat. The
barrier is too low compared to their bulk
energy to lead to significant localization
on one side or the other.
The tunneling suppression factor $\sim (mL)^{2\nu}$ is replaced
by a function which quickly converges to unity when $mL >>1$.
Hence the couplings rapidly converge to
\eqn\heavycouplingsd{\eqalign{
&g^L{}_{L~4D} ={g^L{}_L \over \sqrt{L}} \sqrt{l_1 \over l_1 + l_2}
~~~~~~~~~~~~~~~~~~
g^L{}_{R~4D} = {g^L{}_R \over \sqrt{L}} \sqrt{l_2\over l_1 + l_2}
\cr
&g^R{}_{L~4D} ={g^R{}_L \over \sqrt{L}} \sqrt{l_1 \over l_1 + l_2}
~~~~~~~~~~~~~~~~~~
g^R{}_{R~4D} = {g^R{}_R \over \sqrt{L}} \sqrt{l_2\over l_1 +
l_2}\ .}}
The masses are controlled by the size of the conformal box
inhabited by the bulk modes, $l_1+l_2$:
\eqn\heavymsd{m = {\cal O}(1) {n - \beta \over l_1+l_2} \, ,}
as are the mass splittings, which however also depend inversely
on the mass of the state, as dictated by the SUSY-breaking
boundary condition
\eqn\heavymsplit{
\delta m = {\cal O}(1) {\hat q \over m(l_1 + l_2)} \, .}
This completes the determination of the spectra of bulk modes.

\subsec{Transmission to brane worldvolume fields}

Given the spectrum of the bulk modes and their couplings
to the IR branes, we can compute the induced mass
splitting of scalar superpartners of $4D$ fermions.
The most interesting case is case 2) where the
splitting is suppressed by tunneling effects.
We will compute the splitting for types A) and
B) SUSY breaking.  We will also discuss case 1)
in order to isolate tunneling effects in case 2).

Suppose there are brane-localized scalars and fermions
$\phi^{L,R}{}_{b}$ and $\lambda^{L,R}{}_{b}$, and bulk scalars and fermions
$\phi^{L,R}{}_{B}$ and $\lambda^{L,R}{}_{B}$. Indices $L,R$ refer to the
left- and right-localized states, both in the throat and on the IR brane.
Assume the 4d effective
Lagrangian contains generic
quartic couplings between brane and bulk KK scalars, and Yukawa
couplings between brane and bulk scalars and fermions, of the
form
\eqn\interactions{{\cal L} = \cdots + (g^{I}{}_{J~5D})^2
(\phi^{J}{}_b)^2 (\phi^{I}{}_B)^2 +
g^{I}{}_{J~5D} ~\phi^J{}_{b}~\lambda^J{}_{b}~\lambda^I{}_{B} + \cdots}
Since $[\phi_B] = 3/2$ is the canonical dimension of a 5d scalar
field, one expects
\eqn\gfive{g^I{}_{J~5D} \sim {1\over \sqrt{M_5}}\ .}
Renormalizing the fields to have canonical $4D$ kinetic terms,
we find that the induced $4D$
dimensionless couplings are $g^I{}_{J~4D} \sim {1\over \sqrt{M_5 L}}$.
Our results for both type A) and type B) SUSY breaking
can be summarized as follows, recalling that
$l_1 > l_2$:
\eqn\couplings{\eqalign{
&g^L{}_{L~4D} = {{\cal O}(1) \over \sqrt{M_5 L}}
~~~~~~~~~~~~~~~~~~~~ g^R{}_{R~4D} = {{\cal O}(1) \over \sqrt{M_5 L}}
\sqrt{l_2 \over l_2 + f^2(mL) l_1} \cr
&~~~~~~~~~~~~~~~~~ g^I{}_{J~4D} = {{\cal O}(1) \over \sqrt{M_5 L}}
~f(mL) \sqrt{l_J \over l_I + f^2(mL) l_J} (I\neq J)\ .}}
Here $f$ is the square root of the transmission coefficient
through the barrier, which has the asymptotic forms:
\eqn\fts{
f \simeq \cases{~
(mL)^{2\nu} ~ & ~~~~~ $mL << 1$ \cr
~~ 1 ~ & ~~~~~
$mL >>1 $  \ .\cr }}
The ``off-diagonal" couplings are always much smaller than the
``diagonal" ones because of the tunneling.

The masses of the bulk particles are
\eqn\bmasstr{ m(B,F)_{L,R} = \alpha_{L,R}(n) {n+\beta(B,F,n)_{L,R} \over
l_{1,2} + h(n) l_{2,1}}}
where the coefficients $\alpha$ and $\beta$ depend softly
on $M_0, n$.  The difference in $\beta$ between bosons and fermions
parameterizes SUSY breaking. The function $h$
encodes the behavior of masses and splittings in the
limit when tunneling suppression vanishes. It
vanishes near zero and asymptotes to unity
for large argument, to allow for the transition between, e.g.,
\susmass,\splitleft,\susmassr,\splitright\ to
\heavyms. Its precise form is not important for our conclusions.
Note that for the left-localized modes, we can completely ignore
the term with $h$ since we assume $l_1 > l_2$.

Assume that the fermions living on the
IR brane are protected from mass shifts, e.g. because they belong  in chiral gauge
representations.  The scalars
are not.  A naive estimate of the mass shift
of the scalars on the IR brane would run as follows.
The contribution from one-loop diagrams
involving bulk fields $\phi^I{}_B, \lambda^I{}_B$ can be computed
state-by-state, by first performing the loop integrals over $4$-momenta,
and then subtracting fermionic from bosonic contributions.
This gives approximately
\eqn\mshift{(\delta m^J{}_{I b})^2
\sim (g^I{}_{J~4D})^2  \Big[m^2(B)-m^2(F) \Bigr]
\sim (g^I{}_{J~4D})^2 m_I \delta m_I}
for each KK state in the loop,
where $m_I$ and $\delta m_I$ are the bulk supersymmetric mass and
the boson-fermion bulk mass splitting
induced by SUSY breaking, respectively.  If we insert the
appropriate bulk couplings and splittings and
sum over the entire tower of bulk KK modes,
we get a divergent answer, $(\delta m^J{}_{I b})^2 = \sum_{m \sim 1/l}^{\Lambda_{UV}}
(g^I{}_{J~4D})^2  m_I \delta m_I $, suggesting a cutoff $\Lambda_{UV}$
{\it a priori} distinct from
the UV cutoff used to compute the state-by-state
mass splittings.

This divergence is artificial.
In the 5D picture of the calculation
the SUSY breaking transmission is generated by Feynman
diagrams with bulk particles in the loops and wall particles in
the external legs. The
summation over the $4D$ KK states is the ``integration"
over the momentum in the fifth direction in the
higher-dimensional theory.  In the coordinate representation
this calculation is perfectly well-defined
unless the external particles are at coincident
points. This issue is
handled in the standard way by renormalization.

In the momentum representation
$p_5$ is discretized due to the compactness of
our background in the $z$ direction.  By
first doing the integrals over the $4D$ momenta, we would
then sum over divergent quantities, requiring
independent regularization for both the sum and the integrals.
This is not the correct way to proceed. Instead, we
use the fact that field theory regulates itself away from
coincident points in coordinate representation,
sum over $p_5$ first,
and then integrate over the 4-momenta and renormalize the
final answer, if need be. Since the loop
propagators depend inversely on the square of KK mass,
this sum converges.
Our method is very similar to the regularization
techniques employed in \addv,\adqp,\mirpe.

Now before doing any momentum sums or integrations, the
mass splitting for scalars on the IR brane is:
\eqn\tms{ (\delta
m^J{}_{I b})^2 = {2 \over (2 \pi)^4} \sum_{m} (g^I{}_{J~4D})^2
\Bigl(\int  { d^4p\over p^2 + m^2(B)} - \int {d^4p \over p^2 +
m^2(F)} \Bigr) }
Here the factor of $2/(2\pi)^4$ comes from the symmetry factor
for the Lagrangian \interactions\
and the usual normalization for loop integrals.
The boson and fermion masses are
$m(B,F)=\alpha(n+\beta_{B,F})/l$.  The couplings
and the parameters $\beta_{B,F}, l$
explicitly depend on the masses as explained in \bmasstr.

The induced mass splittings will generally be differences of sums
of the form
\eqn\sms{
S_{I} = {{\cal O}(1) \over 8 \pi^4 M_5L} \sum_{n} F[{\alpha
(n+\beta_I)L\over l})] \int  {d^4p \over p^2 + \alpha^2 (n +
\beta_I)^2/l^2}\ .}
Here $F = g^2 M_5 L$ and its form can be readily read from
\couplings, and $l$ is the denominator in \bmasstr.
A crucial property of $F$ is that it converges to a constant for
large arguments. Indeed, on the real axis $F <l_1/l_2$ for any of
the couplings \couplings. We can now use a standard technique for
the summation of infinite series. Consider the contour integral
\eqn\ci{
\int_C dz ~F({\alpha L z\over l}) ~
{ \cot[\pi(z-\beta_I)]\over p^2 + \alpha^2 z^2/l^2}}
where the contour consists of a large circle oriented
counterclockwise and small clockwise circles around all the poles
of the integrand inside the large circle. Then take the limit
where the radius of the big circle goes to infinity, such that
all the poles are enclosed by it. Since $F$ goes to a constant
for large arguments, the integral along the large circle vanishes
because of the denominator in \ci.  Therefore the integrals
along the small circles add up to zero. They can be readily
computed by the residue theorem. The dependence of the
parameters $\beta$ on the mass is very soft, and we can ignore their
effect on deforming the poles of the cotangent away from the
integers.

We also need to account for the poles
in $F$, which are required because
$F$ is not constant and is bounded for large values
of the argument.
By continuity, any such poles in $F$ are
located outside the circle of radius $1/L$.
They could be imaginary, in which case they would correspond to a state
with a large negative $({\rm mass})^2$.
However, it is very straightforward to understand such poles:
the states heavy enough to cross the barrier classically
are not well-localized, and so the left and the right states mesh together strongly.
If one naively ignores the presence of the
states from one side of the barrier, one is
``reminded" of them precisely by these poles.
Hence, such imaginary poles would just correspond to states localized on the other
side of the barrier, which are added in in the other sum.
They should not be included
in the tunneling calculation twice.
If the poles have a large real part, this means that the
the approximate eigenvalue equations which we have used
to find the spectrum receive corrections, as can be seen from
the form of the
functions \stepcoef\ or \brcoef. Both the
transmission coefficient and the eigenvalue equation depend on those
parameters, and after the heavy masses are appropriately corrected,
these poles are accounted for in the improved summation over the
states in the KK towers.
Therefore we conclude that any such poles of $F$ are spurious and should
be circumnavigated by Wick rotation,
which we have already performed. Hence they will not contribute
to the momentum integrals, and we can ignore them.
With our approximations, the only poles we need to keep are at the integers
along the real axis, and at the complex points $z=\pm i lp/\alpha$.
Therefore:
\eqn\sumev{\eqalign{
\sum_{n} { F(\alpha
(n+\beta_I)L/l) \over p^2 + \alpha^2 (n +
\beta_I)^2/l^2} &= -{l  F(i p L )\over 2i \alpha p}
\cot[\pi({il p \over \alpha } - \beta_I ({i p L }))] \cr
& -  {l  F(-i p L ) \over 2i \alpha p}\cot[\pi({il p \over \alpha }
 + \beta_I ({-i pL }))]\ .
}}
Since $F(x)$ is real on the real axis,
the RHS is obviously real.

Now we isolate the terms in \sumev\ which give the UV divergence
in \sms. Since
\eqn\divis{
\cot[\pi({il p \over \alpha } \pm \beta_I)] =
-i - i {2 \over e^{2\pi lp/\alpha \mp 2i\pi \beta_I}-1}}
we can write $S_I$ as
\eqn\smsonec{\eqalign{
S_{I} =& {{\cal O}(1) \over 8 \pi^4 M_5L} {l\over 2\alpha} \int
{d^4p \over p} \Bigl(F(ipL) + F(-ipL)\Bigr) \cr
&  +{{\cal O}(1) \over 8 \pi^4 M_5L} {l\over 2\alpha} \int
{d^4p \over p} \Bigl(
{2 F(ipL) \over e^{2\pi lp/\alpha - 2i\pi \beta_I (i pL)}-1} +
{2 F(-ipL) \over e^{2\pi lp/\alpha + 2i\pi \beta_I(- i pL)}-1} \Bigr)}}
The first integral in \divis\ is divergent.
However it is independent of the particle quantum numbers, and
so cancels out of \tms.
We will however keep this term in mind, because
it will give the dominant, UV-finite, contribution to the gaugino mass
(see \S4.5).
The second integral is finite, and does not cancel out
of \tms\ because of bulk SUSY breaking.
After the angular integration, the remaining integral is
\eqn\smstwoap{
S^{finite}_{I} = {{\cal O}(1) l\over 4\pi^2 \alpha M_5L}  \int^\infty_0
{dp p^2} \Bigl(
{ F(ipL) \over e^{2\pi lp/\alpha - 2i\pi \beta_I (i pL)}-1} +
{ F(-ipL) \over e^{2\pi lp/\alpha + 2i\pi \beta_I(- i pL)}-1} \Bigr)
}
and it is the leading contribution to \tms.
This integral is obviously
finite due to the exponential momentum cutoff in the
denominator, and is well behaved in the IR due to
the measure factor.  Because of the cutoff the integral is dominated by
the contributions $p \sim 1/l$. We can evaluate it approximately
by the saddle point method, once we have the explicit form of $F$.
Thus the states which determine the mass splittings of the wall
chiral multiplets are the light bulk states. The heavy
states would make it more easily over the barrier, but decoupling reduces
their effect on the transmission. We can now compute
the wall chiral multiplet mass splittings case-by-case.

\medskip
\noindent{\it{Direct SUSY Breaking}}
\medskip

For direct SUSY breaking -- case 1) --
we have $F=1$, $\beta_{B,F} = {\rm const}$ and $l=l_1$ \couplings,\bmasstr.
Substituting \smstwoap\ into \tms, using $\beta_B - \beta_F = \bar q^2 L^2$
\splitleft, we evaluate the integrals by the saddle point method.
We find, using $\sin(2\pi|\beta_B|) \sim {\cal O}(1)$,
\eqn\direct{
(\delta m^L{}_{L b})^2 \sim {\cal O}(1)
{ \bar q^2 L \over  M_5 } {1 \over l^2_1}
\ . }
Due to decoupling, only the lightest modes
contribute significantly to supersymmetry breaking
induced on the left IR brane, even though
couplings are independent of the KK masses. Obviously the mass
splittings induced on the brane are given by the simple exponential scaling
induced by the conformal factor on the brane.
This is similar to \gherpom.

\medskip
\noindent{\it{Tunneling mediation with left bulk SUSY breaking}}
\medskip

Let us turn to tunneling mediation --~case 2) -- starting with
type A) SUSY breaking. We compute the mass
splittings on the right IR brane as induced by loops
of the bulk states.  For left-localized states,
$F = x^{4\nu} l_2/l_1$, $\beta_B - \beta_F = \bar
q^2 L^2$ and $l=l_1$ \splitleft,\couplings. Therefore
\eqn\tunstepl{
(\delta m^R{}_{L b})^2 \sim {\cal O}(1) {\bar q^2 L\over
M_5} {l_2 \over l_1^3} \Bigl({ L \over l_1}\Bigr)^{4\nu}
\ . }
Compared to the direct SUSY breaking \direct,
there is additional dependence on the ratio $l_2/l_1$. This is precisely the
rescaling of dimensionful parameters in the AdS space. The right IR brane
resides closer to the UV brane, and the couplings on it are smaller
than those on the left IR brane precisely by this ratio.
In addition, there is
a significant suppression factor $\Bigl({L \over l_1}\Bigr)^{4\nu}$,
over and above the splitting in
direct transmission \direct. This is an exponential suppression
of the scale of SUSY breaking due to tunneling.

Similarly, for the right-localized bulk states,
$F=l_2/(l_2 + x^{4\nu}l_1)$, $\beta_B - \beta_F = \bar q^2 L^2
x^{4\nu}$ and $l=l_2$ \splitright,\couplings. The mass
splitting is:
\eqn\tunstepr{
(\delta m^R{}_{R b})^2 \sim {\cal O}(1)  {\bar q^2 L\over
M_5}  {1 \over l_2 (l_2 + ( L/l_2)^{4\nu} l_1)}
\Bigl({L \over l_2}\Bigr)^{4\nu}\ .}
There are two extreme limits, depending on which
term dominates the denominator.
In these limits, the mass  splitting transmitted to the right IR brane
by the right-localized bulk multiplets is
\eqn\tunstepfin{
(\delta m_{SM})^2 \sim  {\cal O}(1) {\bar q^2 L\over
M_5}\times \cases{~
 {1 \over l_2^2}
\Bigl({ L \over  l_2}\Bigr)^{4\nu} ~
& ~~~~~ $l_1 < l_2 \Bigl({  l_2\over  L }\Bigr)^{4\nu}$ \cr
~{1 \over l_2 l_1 } ~ & ~~~~~
$l_1 > l_2 \Bigl({  l_2\over  L }\Bigr)^{4\nu}$  \ .\cr }}
The first case is again suppressed relative to \direct\
by the factor of $(L/l)^{4\nu}$.
Comparing eqs. \tunstepl\ and \tunstepfin\ for the relevant
values of parameters we find
that the right-localized states always give a dominant contribution to
the transmission of SUSY breaking to the SM
fields on the right IR brane as long as $l_1 > l_2$.

\eject
\noindent{\it{Tunneling mediation with left brane-localized SUSY breaking}}
\medskip

Finally, we repeat these calculations
for type B) SUSY breaking. For the contribution
of the left-localized states,
$F = x^{4\nu} l_2/l_1$, $\beta_B - \beta_F =
qL^2/(l_1x)$ and $l=l_1$ in \dmassl,\couplings. This
gives
\eqn\tundell{
(\delta m^R{}_{L b})^2 \sim {\cal O}(1) { q \over
M_5} {l_2 \over l_1^3} \Bigl({ L \over l_1}\Bigr)^{4\nu} \ .}
Again, the factor of
$\Bigl({ L \over  l_1}\Bigr)^{4\nu} $ reflects
tunneling, while the power $l_2/l_1$ represents
the usual coupling rescaling.

For the right-localized states,
$F=l_2/(l_2 + x^{4\nu}l_1)$,
$\beta_B - \beta_F = x^{4\nu}qL^2/(l_1x)$
and $l=l_2$ in \dmassr,\couplings, and therefore
\eqn\tundelr{
(\delta m^R{}_{R b})^2 \sim {\cal O}(1)  { q \over
M_5} {1 \over l_2(l_2 + ({L / l_2})^{4\nu}l_1) }
\Bigl({ L \over l_2}\Bigr)^{4\nu} \ .}
As for the transmission of left bulk SUSY breaking, we again focus on
two extreme cases,
$l_1 < l_2 ({ l_2\over  L})^{4\nu}$ and $l_1 > l_2
( { l_2\over L})^{4\nu}$.
In either case, we can see that the right-localized modes give dominant contribution to the
splittings of the right brane multiplets.
The transmitted mass
splittings on the right IR brane from the SUSY breaking which
occurs on the left IR brane is
\eqn\tundelfin{
(\delta m_{SM})^2 \sim {\cal O}(1) { q \over
M_5} \times \cases{~
{1 \over l_2^2 }
\Bigl({ L \over l_2}\Bigr)^{4\nu}  ~ & ~~~~~
$l_1 < ({ l_2\over  L})^{4\nu} l_2$ \cr
~ {1 \over l_1 l_2} ~ & ~~~~~
$l_1 > ({ l_2\over L})^{4\nu} l_2$  \cr }}
The main contribution to $\delta m_{SM}$
is induced by the modes living on the right side.
Just like \tunstepfin, $\delta m_{SM}$ is dramatically suppressed compared to
case 1).  The calculations
verify the simple intuition that, to transmit the
SUSY breaking from the left throat to the right throat, the
modes pay an exponential price to tunnel
through the potential barrier at the UV
brane.

\subsec{R-symmetry Breaking and Gaugino Masses}

In the previous sections we have assumed that SUSY breaking
is mediated by the bulk scalars, while bulk fermions were protected
from the breaking effects. However, in order to generate
gaugino masses, bulk fermions must also couple
to SUSY-breaking physics.
Only through their exchange is it possible to break
R-symmetry, which is necessary for gaugino mass generation.

We will consider in detail a scenario where SUSY breaking is
confined to the left IR brane. One could, in principle,
break SUSY in the entire left throat.
However, the combination of
the Dirac mass term (needed to avoid
tachyons among the scalar superpartners) and the Majorana mass term
(necessary to break R-symmetry and generate gaugino masses)
for bulk fermions
complicates the calculation considerably.

Our model consists of two chiral
bulk fermions, $\zeta$ (undotted, two-component
spinor -- \cf\ \wessbagg\ for notation) and
$\chi$ (dotted spinor) minimally coupled to gravity and coupled to
each other by a bulk Dirac mass $M$,
and left-brane-localized Majorana terms induced by SUSY breaking.
In this section we will work with
the metric signature $(+----)$.

A simple action for the fermions is:
\eqn\fermact{\eqalign{
S_F &= \int d^5 x \sqrt{g_5} \Bigl({i \over 2} \bar \Psi \Gamma^A
\tp_A \Psi + {1\over 8} \omega_{bc.A}
\bar \Psi \{\Gamma^A, \sigma^{bc}\} \Psi - M\bar \Psi \Psi \Bigr)
\cr
&+ \int d^4 x \sqrt{g_4}\Bigl( q_1 [ \zeta^T i\sigma^2
\zeta -  \zeta^\dagger i\sigma^2
\zeta^*] - q_2 [ \chi^T i\sigma^2
\chi -  \chi^\dagger i\sigma^2
\chi^*] \Bigr) \ .}}
Here $\Psi^T = (\zeta, \chi)$, and
\eqn\twosder{
    A\ \tp\ B = A\p B - (\p A) B\ .
}
Also, $\gamma^A = (\gamma^\mu, {\rm diag}(-i,i))$
are $5D$ Dirac matrices, $\Gamma^A = E_B{}^A \gamma^B$, and
$\sigma^{ab} = {i \over 2} [\gamma^a, \gamma^b]$.
$g^{5D}_{AB} = \eta_{CD} e^C{}_A e^D{}_B$ defines both the funfbein
$e^A{}_B$ and its inverse $E_B{}^A$, $e^A{}_B E_C{}^B =
\delta^A{}_B$.
Hermitean conjugation is denoted by
$\dagger$, complex conjugation by $*$ and transposition by $T$.
Dirac conjugates are defined on the tangent
space $\bar \Psi = \Psi^\dagger \gamma^0$. Finally, will will use
the spinor representation of the Dirac matrices:
\eqn\dirac{
\gamma^\mu = \left(\matrix{~0~&~\sigma^\mu~\cr
 \bar \sigma^\mu~&~0~\cr}\right)}
and $\sigma^\mu = {\rm diag}(1,\sigma^k)$,
$\bar \sigma^\mu = {\rm diag}(1,-\sigma^k)$ and $\sigma^k$ are
the usual Pauli matrices.

Before proceeding we note that because the Majorana terms
live on the left IR brane, they don't need to be the symplectic
Majorana masses characteristic for bulk $5D$ fermions;
our background breaks invariance under the AdS isometry group.
We will see below
that one of the mass parameters $q_i$ must vanish identically.
Furthermore, because the $5D$ fermions
dimension $2$, the coefficients $q_1, q_2$
of the Majorana terms on the left IR brane are
dimensionless. This effects the
dependence of the induced mass splitting on the exponentials
that emerge from the width of the barrier between the
throats, leading to different SUSY-breaking
behavior than that discussed previously
for bosons. The $4D$ scale which emerges in the effective $4D$
description of these modes is induced by the normalization of the
modes and their overlap with the left IR brane, and can therefore
be small.

We emphasize that while the IR branes are endowed with
orbifold boundary conditions, the UV brane is {\bf not}. For fermions
this affects physics in a crucial way. An immediate consequence is that
the bulk Dirac mass parameter $M$ is continuous across the UV brane,
$M(-z)=M(z)$. As a result the bulk parity operator acting on fermions is not
the same on all the branes.
We will see below that parity inversion about
the UV brane interchanges the chiral spinors
while it does not around IR orbifolds.
If we take the realization of parity on the
UV brane as ``natural", then the choice
of parity assignments on different IR branes can lead to
twisted bulk fermions, breaking left-right symmetry about the UV brane.
Such bulk fermions are not degenerate even in
the supersymmetric limit, despite the apparent symmetry of \metricL.
We also note that these boundary conditions on the UV brane
would be compatible with additional UV brane-localized supersymmetric fermion
mass terms, analogous to the $\sim \delta$-function supersymmetric terms
in \bessel\ which we have ignored. We will ignore such terms here
too.

Note that the bosons of \S4.3 and fermions
treated here are not supersymmetric partners of each other.
The correct sfermions would have to satisfy different boundary
conditions from those we have chosen in \S4.3.
We will ignore a detailed treatment of the latter here
and merely point out
the relations later on. Our analysis is precise for the
case when SUSY breaking on the left IR brane is completely
confined to the fermion sector.

Now we wish to find the equations of motion for the fermions.
Because the
background \metricL\ is conformally flat, $\omega_{bc.A} \bar \Psi
\{\Gamma^A,\sigma^{bc}\} \Psi = 0$ identically. Substituting the
ansatz \metricL\ into \fermact, after straightforward algebra we find
\eqn\fermactred{\eqalign{
S_F &= \int d^4x dz \Bigl\{ a^4 \Bigl[
\chi^\dagger i \sigma^\mu \partial_\mu \chi
+ \zeta^\dagger i \bar \sigma^\mu \partial_\mu \zeta
+ \chi^\dagger ({1\over 2} \tp_z \zeta
+ 2 {a' \over a} \zeta)
- \zeta^\dagger ({1\over 2} \tp_z \chi
+ 2{a'\over a} \chi)\Bigr] \cr
& - Ma^5(\chi^\dagger \zeta + \zeta^\dagger
\chi)+ \delta(z+l_1) a^4 \Bigl( q_1 [ \zeta^T i\sigma^2
\zeta -  \zeta^\dagger i\sigma^2
\zeta^*] - q_2 [ \chi^T i\sigma^2
\chi -  \chi^\dagger i\sigma^2
\chi^*] \Bigr)\Bigr\}\ . }}
Defining the effective $4D$ theory fermions by the
change of variables $Z = a^2 \zeta$, $X=a^2 \chi$ removes the
cross terms $\sim a' \Psi'$ from the action, so that:
\eqn\fermacred{\eqalign{
S_F &= \int d^4x dz \Bigl\{
X^\dagger i \sigma^\mu \partial_\mu X
+ Z^\dagger i \bar \sigma^\mu \partial_\mu Z
+ {1\over 2} X^\dagger  \tp_z Z
- {1\over 2} Z^\dagger \tp_z^ X
 - Ma(X^\dagger Z + Z^\dagger X)\cr
&~~~~~~~~~~~~~~~~ + \delta(z+l_1) \Bigl( q_1 [ Z^T i\sigma^2
Z -  Z^\dagger i\sigma^2
Z^*] - q_2 [ X^T i\sigma^2
X -  X^\dagger i\sigma^2
X^*] \Bigr)\Bigr\} \ .}}
The kinetic terms are now independent of the warp
factor, which only enters through the bulk
Dirac mass term. This reflects the
conformal symmetry of a massless fermion; the symmetry
is broken only by the mass terms.

The fermion equations of motion are:
\eqn\fereqs{\eqalign{
&i\bar \sigma^\mu \p_\mu
    Z - X' = Ma X - 2 q_2 \delta(z+l_1) i\sigma^2 Z^* \cr
&i \sigma^\mu \p_\mu
    X + Z' = Ma Z + 2 q_1 \delta(z+l_1) i\sigma^2 X^* \ .}}
These equations can be cast in 4-component form by
introducing
\eqn\matrices{
\Sigma =  \left(\matrix{~Z~\cr
~X~\cr}\right) ~~~~~~~~~
D =  \left(\matrix{~\partial_z~&~i\sigma^\mu \partial_\mu~\cr
i \bar \sigma^\mu \partial_\mu~&~-\partial_z~\cr}\right)
~~~~~~~~~
Q =  \left(\matrix{~0~&~2q_1 i \sigma^2~\cr
-2q_2 i \sigma^2 ~&~0~\cr}\right) \ ,}
whereupon \fereqs\ become
\eqn\fereq{
D \Sigma = Ma \Sigma + \delta(z+l_1) Q \Sigma^*\ . }
The second term on the RHS seems to complicate matters,
but it is easy to reinterpret as a shift in the
boundary condition on $\Sigma$.

To find this shift, we must first address the
$z$-parity of the fermion on the orbifold branes.
The bulk Dirac mass term must be odd on each orbifold
$z_o$, and the fermion must satisfy
\eqn\parity{ i\gamma^5
\Sigma(z_o-\epsilon) = \pm \Sigma(z_o +\epsilon)}
The signs can be chosen independently on each orbifold and
different choices represent different superselection sectors.
The four sectors can be classified by the $({\rm left,right})$
pairs $(++)$, $(+-)$, $(-+)$ and $(--)$. For
definiteness in what follows we will consider $(++)$ and $(+-)$ cases.
We will see that the former corresponds to twisted fermions in the
bulk. The other two combinations are similar to these two with the
interchange $Z \leftrightarrow X$.
Hence if the parity assignment on an orbifold brane is $e^{in\pi}$,
\eqn\parcomp{ Z(z_o-\epsilon) =
e^{in\pi} Z(z_o +\epsilon) ~~~~~~~~~~~~~
X(z_o-\epsilon) = - e^{in\pi} X(z_o+\epsilon)\ .}
Then by Gaussian pillbox integration of \fereq\ around
the left IR (+) brane we obtain fermion boundary conditions there:
\eqn\bcferir{q_1 X(-l_1) = 0
~~~~~~~~~~~~~~~~~~~~~~ X(-l_1) = q_2 i \sigma^2 Z^*(-l_1)\ . }
If both $q_1, q_2 \ne 0$, then $Z(-l_1)=X(-l_1)=0$,
which has no nonzero solutions.
We will therefore choose $q_1=0$.
Near the right IR brane there are no SUSY-breaking terms in the action,
implying $X(l_2)=0$
for the $(++)$ case, and $Z(l_2)=0$ in the $(+-)$ case. On the UV brane we
demand continuity of the wavefunction,
since we do not impose orbifold
boundary conditions there. The
equations of motion and boundary conditions become:
\eqn\bvpfer{\eqalign{
& D \Sigma = M a \Sigma \cr
& X(-l_1) = q_2 i \sigma^2 Z^*(-l_1)
\cr & Z(0_+) = Z(0_-)
~~~~~~~~~~~~~~~~~~~~~~~~~~~ X(0_+) = X(0_-) \cr
&X(l_2) = 0 ~~~~~ (++)~{\rm case}  ~~~~~~~~~~~~~~~~
Z(l_2) = 0 ~~~~~ (+-)~{\rm case}
}}
The appearance of the complex
conjugate of the fermion field in
the boundary condition of the left IR brane relates
the positive and negative energy solutions there.
We now turn to determining the spectrum of its solutions.

First, the system \bvpfer\ does not
admit any $4D$ zero modes. Such modes
would satisfy $\gamma^\mu \partial_\mu \Sigma =0$. The bulk equation
would then reduce to the system
\eqn\ferzero{
Z' = Ma Z ~~~~~~~~~~~~~~ X' = - Ma X\ .}
Defining $w = m(|z|+ L)$ \newvar, $\nu = ML$
(note the difference between this and \nudef),
the solutions are
\eqn\solferzer{Z = \cases{~ Z_{0 ~L} ~w^{-\nu} ~ & ~~~~~ $z < 0$ \cr
~ Z_{0 ~R} ~w^\nu ~ & ~~~~~ $z > 0$  \cr } ~~~~~~~~~~~~~~~~
X = \cases{~ X_{0 ~L} ~w^\nu ~ & ~~~~~ $z < 0$ \cr
~ X_{0 ~R} ~w^{-\nu}~ & ~~~~~ $z > 0$ \cr}}
where $Z_{0~L,R}, X_{0~L,R}$ are constant chiral spinors.
Note the manifestation of the action of
parity about the UV brane: under $z \leftrightarrow - z$,
we must interchange $Z \leftrightarrow X$.
This should be contrasted with the
orbifold boundary conditions \parcomp. This also clearly shows why
$(++)$ case corresponds to twisted fermions and $(+-)$ to
normal ones. If we impose boundary conditions in the former case,
we have different functions on different sides ($\sim
w^\nu$ to the left and $\sim w^{-\nu}$ to the right). In the
latter case, the boundary conditions are imposed on the same
function ($\sim w^\nu$).

Inserting \solferzer\
into \bvpfer,
we find that the only solutions
are $Z_{0~L} = Z_{0~R} = X_{0~L} = X_{0~R} = 0$.
This differs from the single-throat
situation in \ygmn. There, both the UV and
IR branes were endowed with orbifold boundary
conditions. Hence one of the chiral spinors
always satisfied the boundary conditions trivially,
allowing for an arbitrary constant
profile in the bulk. This gave rise to a fermion zero mode in the effective
$4D$ theory.

Now we consider massive $4D$ modes. The linear equation
\bvpfer\ can be diagonalized by squaring the operator $D$, as
$D^2 = \partial^2_z - \partial^2_4$ is diagonal in spinor indices.
$\Sigma$ satisfies
$(D^2 - M^2 a^2)\Sigma = Ma' ~i\gamma^5 \Sigma$, which is diagonal
in the spinor representation. Set $ \Sigma = \Sigma(z) \exp(-ip\cdot x)$,
so that $\partial^2_4 \Sigma = - m^2 \Sigma$; and set $w = m(|z|+L)$
and $\nu = ML$.  We find:
\eqn\compeqs{\eqalign{
{d^2 Z \over dw^2} + Z = &\cases{~ {\nu(\nu+1) \over w^2} Z ~ & ~~~~~ $z < 0$ \cr
~ {\nu(\nu-1) \over w^2 }Z ~ & ~~~~~ $z > 0$  \cr } \cr
{d^2 X \over dw^2} + X = &\cases{~ {\nu(\nu-1) \over w^2 }X ~ & ~~~~~ $z < 0$ \cr
~ {\nu(\nu+1) \over w^2 }X ~ & ~~~~~ $z > 0$  \cr }}}
These are just Bessel equations.
Before proceeding let us contrast this case to that when
UV brane is an orbifold \ygmn.
In the latter case $M$, and
therefore $\nu = ML$, changes sign across the UV brane,
ensuring that the potential on the two sides of the UV brane
does not change. In our problem the Dirac mass is
continuous across the UV brane; so the bulk potential for the chiral
spinors changes across the brane, as is manifest in \compeqs.
Therefore the solutions on the right side are obtained from those on
the left by $z \leftrightarrow -z$, $Z \leftrightarrow X$, so
we need only explicitly solve for $z < 0$.
As a result the $(++)$ boundary conditions break
the parity symmetry of \metricL, since they twist the fermion modes,
while the $(+-)$ boundary conditions are parity-invariant.

We also note that the form of \compeqs\ shows that there have to
be two sfermions with bulk masses related to the bulk fermion
masses by $\sqrt{4+M^2_B L^2} = M_F L \pm 1/2$; only in this
case would the bulk equations for both bosons and fermions admit
the same spectrum. Furthermore, the boson boundary conditions must be
chosen appropriately so that they reproduce the mass
spectrum of the fermions.
In what follows we will therefore assume that the spectrum of
sfermions coincides with that of fermions in the SUSY limit, and
that SUSY is broken exclusively by the $\sim q_2$ term.

Now, the solutions
for $z < 0$ are:
\eqn\leftferso{\eqalign{
& Z = \sqrt{w} \Bigl(A J_{\nu+1/2}(w) + B J_{-\nu - 1/2}(w) \Bigr) \cr
& X = \sqrt{w} \Bigl(C J_{\nu-1/2}(w) + D J_{-\nu + 1/2}(w) \Bigr)\ . }}
The constant spinors $A,B,C,D$ are not all independent
due to the equation of motion in \bvpfer\ for $\Sigma$.
In the rest frame of the massive modes, \bvpfer\ becomes
\eqn\linres{
\left(\matrix{~-\partial_w~&~1~\cr
1~&~\partial_w~\cr}\right) \Sigma = {\nu \over w} \Sigma\ .}
Upon substituting \leftferso, we find
$C=A, D= -B$.
Using the map $z \leftrightarrow -z$, $Z \leftrightarrow X$,
the bulk fermion wavefunctions in four-component notation are
\eqn\bulkfersol{\eqalign{
& \Sigma_L = \sqrt{w} \left(\matrix{
~A_L J_{\nu+1/2}(w) + B_L J_{-\nu - 1/2}(w)~ \cr
~A_L J_{\nu-1/2}(w) - B_L J_{-\nu + 1/2}(w)~ \cr} \right) ~~~~~~~~ z < 0\cr
& \Sigma_R = \sqrt{w} \left(\matrix{
~A_R J_{\nu-1/2}(w) - B_R J_{-\nu + 1/2}(w)~ \cr
~A_R J_{\nu+1/2}(w) + B_R J_{-\nu - 1/2}(w)~ \cr} \right) ~~~~~~~~ z > 0
}}
where $A_{L,R}, B_{L,R}$ are constant, momentum-dependent
$2$-component spinors encoding $4D$ fermion helicities
(and are not $c$-numbers, in contrast to \ygmn). In an arbitrary frame,
the basis spinors are given by Lorentz boosts of these.

To determine the spectrum we substitute the solution \bulkfersol\ into the
boundary conditions specified in \bvpfer.
We will use the shorthand notation $j_{\alpha,k} = J_\alpha(m(l_k+L))$,
$J_\alpha = J_\alpha(mL)$ and
$J_\alpha(-m(l_k+L))= \exp(2\pi i \alpha) ~j_{\alpha,k}$.
The analysis of the boundary conditions
on the right IR brane and on the UV brane is
straightforward; analysis of the boundary condition on the left IR brane is
more subtle. As the latter involves $i \sigma^2 Z^*(-l_1)$,
in the momentum basis the mode which appears under
complex conjugation is the
negative energy mode.
Defining the phase $\theta = 2\pi  \nu$, we find:
\eqn\irferbc{\eqalign{
&A_L j_{\nu-1/2,1} - B_L j_{-\nu+1/2,1} =
- q_2 i\sigma^2 A_{L~c}^* e^{-i \theta} j_{\nu+1/2,1}
- q_2 i\sigma^2 B_{L~c}^* e^{ i \theta} j_{-\nu-1/2,1} \cr
&A_{L~c}  ~e^{i \theta}j_{\nu-1/2,1} +
B_{L~c}  ~e^{- i \theta} j_{-\nu+1/2,1} =
- q_2 i\sigma^2 A_{L}^* j_{\nu+1/2,1}
- q_2 i\sigma^2 B_{L}^* j_{-\nu-1/2,1}\ .}}
The remaining two boundary conditions then give
\eqn\simplebc{\eqalign{
 B_R = - {j_{\nu+1/2,2} \over j_{-\nu-1/2,2}} A_R ~~~~(++)~ {\rm case}
~~~~~~
&~~~ B_R = {j_{\nu-1/2,2} \over j_{-\nu+1/2,2}} A_R ~~~~(+-)~ {\rm case}\cr
A_L\left(\matrix{ J_{\nu+1/2} \cr
 J_{\nu-1/2} \cr} \right)
 + B_L \left(\matrix{J_{-\nu - 1/2}\cr
 - J_{-\nu + 1/2} \cr} \right)
~= & ~~A_R \left(\matrix{ J_{\nu-1/2} \cr
J_{\nu+1/2} \cr} \right)
 + B_R \left(\matrix{-J_{-\nu + 1/2}\cr
 J_{-\nu - 1/2} \cr} \right)\ ,
}}
and two more equations where $A^c_{L,R}, B^c_{L,R}$ replace
$A_{L,R}, B_{L,R}$, and $m \rightarrow - m$. Here the superscript
$c$ denotes the negative energy mode components, or charge
conjugates.
Note that if $q_2$ were zero, the $(+-)$ boundary conditions
involve the same functions of $ml$ on both the left and right
sides, while $(++)$ do not, again showing that the latter are
twisted fermions. In either case there are
eight boundary conditions for nine integration parameters:
eight constant spinors $A_{L,R}, B_{L,R}, A_{L,R~c}, B_{L,R~c}$ and the $4D$
mass $m$. One of the spinors is fixed by the overall normalization;
for the remaining seven spinors the mass must be selected to solve
the eight boundary conditions, wherefore the spectrum is discrete.

We can solve this system of equations explicitly, expressing all
the spinors in terms of $A_L$ which we choose by overall
normalization for convenience.  The masses are determined by the
roots of the equation:
\eqn\eigfer{\eqalign{ &\eta_1 \Bigl(j_{\nu-1/2,1} \mp q_2
j_{\nu+1/2,1}\Bigr) = \eta_2 \Bigl(j_{-\nu+1/2,1} \pm q_2 j_{-\nu-1/2,1}
\Bigr) \cr
&\eta_1 =  J_{-\nu+1/2} J_{\nu-1/2} + J_{\nu+1/2}
J_{-\nu-1/2} \pm (J^2_{-\nu+1/2} - J^2_{-\nu-1/2} ) {j_{\nu\pm
1/2,2} \over j_{-\nu \mp 1/2,2}} \cr
&\eta_2 = J^2_{\nu-1/2} -
J^2_{\nu+1/2} \pm (J_{-\nu+1/2} J_{\nu-1/2} + J_{\nu+1/2}
J_{-\nu-1/2}) {j_{\nu \pm 1/2,2} \over j_{-\nu \mp 1/2,2}}\ .
}}
In the first of these equations the $\pm$ sign changes for different
chiralities, whereas in the latter two equations it changes with
the boundary condition choices, being + for the $(++)$ case and
$-$ for the $(+-)$ case.

To extract the physical properties of the spectrum from this
quagmire, we first replace $L$ by $\zeta L$ everywhere in these
equations and then take a double limit $q_2 = 0$, $\zeta
\rightarrow 0$, keeping $ml_1, ml_2$ fixed. This corresponds to removing
SUSY breaking terms from the IR brane and
(following the discussion at the beginning of Sec. 2)
removing the UV brane, separating the two throats by an infinite distance.
In this limit, the spectrum splits into separate towers of
left- and right-localized modes.  These modes
modes will not be degenerate, since \metricL\ is not symmetric.
Using the small argument expansion of Bessel functions
$J_\alpha(m \delta L) \sim (m \zeta L)^\alpha$, we see that to
leading order $\eta_1 \sim (m \zeta L)^{- 2\nu -1}$ is divergent,
while $\eta_2 \sim {\cal O}(1)$. Hence analyticity requires that
the divergent part of the $\eta_1$ term
must vanish, leading to the
eigenvalue equations $J_{\nu-1/2}(ml_1) J_{\nu \pm
1/2}(ml_2) = 0$, where we ignore terms ${\cal O}(L/l_1, L/l_2) << 1$.
Therefore, the bulk fermion states split into two
towers with masses given by the roots of
\eqn\susyferspec{\eqalign{ &J_{\nu-1/2}(ml_1) = 0
~~~~~~~~~~~~~~~~~~~~~~~~~~ {\rm left-localized ~states}\cr
&J_{\nu\pm 1/2}(ml_2) = 0 ~~~~~~~~~~~~~~~~~~~~~~~~~~ {\rm
right-localized ~states}
}}
Let us check these conditions.
Consider the solution \bulkfersol. The $\sim B_{L,R}$
terms diverge in the limit $\zeta \to 0$;
normalizability of the solution requires
$B_{L,R}=0$ in that limit. The boundary condition on the UV brane
is then satisfied. Moreover, when SUSY is unbroken,
$q_2=0$ and the boundary conditions on either IR brane are $X=0$.
On the left this gives precisely $J_{\nu-1/2}(ml_1)=0$ while on the
right for the $(+\pm)$ cases it gives $J_{\nu \pm 1/2}(ml_2)=0$.

When supersymmetry breaking is turned on and $\zeta=1$,
the masses of these states get contributions from
both supersymmetric UV effects and from
SUSY-breaking effects. As we have explained above,
we only wish to compute the SUSY-breaking
induced mass shifts to leading order in the (dimensionless) SUSY breaking
parameter $q_2$.  For left-localized states,
Eq. \eigfer\ implies that, to leading
order, the SUSY-breaking induced splittings are determined by the
roots of \eqn\susbrlf{j_{\nu-1/2,1} = \pm q_2 j_{\nu+1/2,1}}
Therefore, using the large-argument form of the Bessel functions
$J_{\nu-1/2}(x) \sim \sqrt{2 \over \pi x} \cos(x-\nu
\pi/2)$, we find that \susbrlf\ reduces to $\cos[m(l_1+L) -
\nu\pi/2] = \pm {\cal O}(1) q_2$. Thus the SUSY masses of light states
are \eqn\susmasfer{m \sim [{(\nu -1)\pi \over 2}
    + n\pi] {1 \over l_1}\ ,}
and the SUSY-breaking mass splittings are
\eqn\susbrlfm{\delta m_L =
\pm {\cal O}(1) {q_2 \over l_1} \ ,}
for both the $(++)$ and $(+-)$ cases.
Therefore SUSY breaking lifts the degeneracy between chiral
fermions.

To compute the transmission of SUSY breaking to the right wall,
we need to determine the couplings. Naively, the fermion boundary
conditions \bvpfer\ might suggest that the $(++)$ case does
not give an
efficient SUSY breaking mechanism, because $X=0$ on the right wall
and there is no manifest Majorana mass for $Z$ in that case.
However the $z$-derivative terms in \fermacred\
give Majorana masses upon Kaluza-Klein reduction,
with the same order of magnitude as the manifest $Z$ Majorana masses.
Then from \irferbc,\simplebc\ we can deduce that for these states
$A_R, B_R, B_L \sim (mL)^{2\nu} A_L$ and so these modes are
dominated by the $\sim A_L$ component in \bulkfersol\ throughout
most of the bulk geometry. Because $l_1 > l_2$,
the proper normalization of the
wavefunction is achieved by setting $A_L \sim 1/\sqrt{l_1}$
paralleling the case of bosons from \S4.3. Therefore $A_R \sim
(mL)^{2\nu}/\sqrt{l_1}$. Recalling the form of the fermionic interaction
Lagrangian \interactions\ and the conformal scalings of the fields
$\phi_{4D} = a \phi_b$, $\lambda_{4D~b} = a^{3/2} \lambda_b$
and $\lambda_{4D~B} = a^2 \lambda_B = a^2 \Sigma$, the effective
$4D$ couplings are the same as that for the bosons. If
$g^I{}_J = {\cal O}(1)/\sqrt{M_5}$, the
left-localized fermion couplings to the
fields on the right IR brane are
\eqn\fercoups{
\eqalign{
& g^L{}_{L~4D} = {{\cal O}(1) \over \sqrt{M_5L}} \cr
& g^L{}_{R~4D} = {{\cal O}(1)\over \sqrt{M_5L}}
\sqrt{l_2\over l_1} (mL)^{2\nu}\ .}}

For heavy fermions, the tunneling suppression weakens, and
eventually disappears for the states with $mL >1$. The masses and
the couplings of these states behave essentially as those for
bosons, in eqs. \heavyms,\heavycouplings.

Now we look at the right-localized states and compute their mass
splitting and $4D$ effective theory couplings.
For these states,
$A_L,B_L, B_R \sim (mL)^{2\nu} A_R$ and so
the correct normalization requires
$A_R \sim 1/\sqrt{l_2 + (mL)^{4\nu}l_1}$. The eigenvalue equation becomes
\eqn\rightfereig{ j_{\nu \pm 1/2,2} = {\cal O}(1) (mL)^{2\nu+1} \mp
q_2 {\cal O}(1) (mL)^{4\nu}\ ,}
and to leading order the supersymmeric masses and the
SUSY-breaking induced mass splittings
are
\eqn\susmasferr{m \sim [{(\nu -1)\pi \over 2} + n\pi] {1 \over l_2}}
\eqn\rigferms{
\delta m = \pm {\cal O}(1) {q_2 \over l_2} (mL)^{4\nu} \ .}
The effective $4D$ couplings of the right-localized fermions
are
$$g^R{}_{i~4D} \sim g^R{}_i (A_i/A_R) /\sqrt{a l} \sim
g^R{}_i (A_i/A_R) /\sqrt{L}\ ,$$ so that
\eqn\rfercoups{
\eqalign{
& g^R{}_{L~4D} = {{\cal O}(1) \over \sqrt{M_5L}} (mL)^{2\nu} \sqrt{l_1 \over
l_2 + (mL)^{4\nu}l_1}  \cr
& g^R{}_{R~4D} = {{\cal O}(1)\over \sqrt{M_5L}} \sqrt{l_2 \over
l_2 + (mL)^{4\nu}l_1}\ .
}}
Now we are ready to compute the SUSY breaking transmission to the
fields localized on the right IR brane.

As a warmup, we consider the radiatively-induced scalar masses.
We begin with terms $\sim m_s^2 \phi_b^\dagger \phi_b$, for any brane scalars
$\phi_b$ that may have direct couplings to the split bulk
multiplets. To compute the masses, we again use the formulas \tms\ and \smstwoap\
derived in \S4.4. Using \susbrlfm, \fercoups, \rigferms\ and \rfercoups,
we see that for the left-localized modes, $F= x^{4\nu}l_2/l_1 $, $\beta_B - \beta_F
= q_2 $ and $l = l_1$. Therefore the scalar masses induced by the
left-localized modes are
\eqn\squarkmassl{
m^2_{s} \sim {\cal O}(1) {q_2 \over
M_5L} {l_2 \over l_1^3} \Bigl({L \over  l_1}\Bigr)^{4\nu}  }
For the right-localized modes, $F=l_2/(l_2 + x^{4\nu}l_1)$,
$\beta_B - \beta_F = q_2 x^{4\nu}$ and $l=l_2$, and so
\eqn\squarkmassr{
m_{s}^2 \sim {\cal O}(1) {q_2\over
M_5L}  {1 \over l_2 (l_2 + (L/ l_2)^{4\nu} l_1)}
\Bigl({ L \over l_2}\Bigr)^{4\nu}\ .}
Here again if $l_1 < l_2 \Bigl({ l_2\over L }\Bigr)^{4\nu}$,
the limit is
\eqn\sqmrf{ m_{s}^2 \sim
{\cal O}(1) { q_2 \over
M_5L}  {1 \over l_2^2}
\Bigl({ L \over  l_2}\Bigr)^{4\nu}\ .}
When $l_1 > l_2 \Bigl({  l_2\over  L }\Bigr)^{4\nu}$
the scalar masses are
\eqn\sqfinsm{
m_{s}^2 \sim {\cal O}(1) {q_2\over
M_5L}  {1 \over l_2 l_1 } \ .}
Hence again the right-localized states give a dominant contribution to
the scalar masses as long as $l_1 > l_2$. Similar SUSY-breaking
mass splittings are also induced for
any other right IR-brane chiral supermultiplet,
in particular for an adjoint multiplet.
Because these terms contain bulk fermions with Majorana
mass in the loop, they break $R$-symmetry.

We further consider SUSY breaking-induced masses of the form
$m^2_A \phi_b^2$, which violate particle number conservation. Such
terms arise for example in theories which contain boundary
interactions
\eqn\lagr{ {\cal L}_{int} = g^2_5
\Phi_b^2(x) [\Phi_B^\dagger(x,z=l_2)]^2 + \mu
\Phi_B^2(x,z=-l_1) \, ,}
where the first term denotes the interaction
between the right brane scalars and the bulk scalars while the
second term stands for the SUSY-breaking left brane mass term for
the bulk scalars.
Note that the quartic interaction term can arise from
the effective $4D$ superpotential
\eqn\suppot{ W = W_{SSM} +
g{\cal F}_b^3 +g {\cal F}_b \sum_m {\cal F}_B^2 \, ,}
where $\phi_b$ and $\phi_B$ are the superfields containing
$\Phi_b$ and $\Phi_B$.
To proceed, we need to reduce \lagr\ to 4d, and determine the
terms it
gives rise to in the effective 4d theory. With the usual rules
for dimensional rescaling for the KK modes, and the normalizations
of bulk wavefunctions which specify the overlap,
we find that the effective 4d theory is
\eqn\efflag{
{\cal L}_{eff} = g_4^2 \phi_b^2 [\phi^\dagger]^2_B + \delta m^2_B \phi_B^2
\, , }
where the couplings and mass insertions for the left localized
modes are $g^2_{4D} \sim {\l_2 \over M_5Ll_1} (mL)^{4\nu}$ and
$\delta m_B^2 \sim 1/l_1^2$ for light modes and vanishingly
small $\delta M_B^2$ for heavy modes. For right-localized modes,
$g_{4D}^2 \sim {1\over M_5L}$ while
$\delta m^2_B \sim {q_2 \over l_1(l_2 + (mL)^{4\nu}l_1)}
(mL)^{4\nu}$ for light modes and again vanishingly small $\delta
m_B^2$ for heavy modes.
These vertices give rise to
a one-loop Feynman diagram with $\phi_b$ as external legs and
$\phi_B$ loop, and one mass insertion $\sim m^2_B$, which generates
the one-loop mass term $\sim m_A^2 \phi^2_b$ where \eqn\nummass{
m_A^2 \sim  {\cal O}(1)\sum_{m} g^2 \delta m^2_B \int {d^4p \over
(p^2 + m_B)^2} \, .} It is now clear that the right-localized modes
dominate the contribution to \nummass. We can now evaluate the contribution of the
left-localized modes by our summation technique, noticing that by
defining $F = g^2 \delta m^2_B M_5L$ and $(p^2 + m_B)^{-2} = -
p^{-2}\partial_s (sp^2 + m_B)^{-1}|_{s=1}$ we get \eqn\numas{m_A^2
\sim  -{{\cal O}(1) \over M_5L}
\partial_s  \int {d^4p \over  p^{2}} \sum_{m}{ F  \over sp^2 + m_B}|_{s=1} \, .}
After summing over the KK masses and performing the angular
integrals, we obtain to leading order
\eqn\numasn{m_A^2
\sim  {{\cal O}(1) l_2 \over M_5L}
 \int dp \Bigl(F(i\sqrt{s}pL) + F(-i \sqrt{s} pL)\Bigr)  \, ,}
where we have used eqs. \sumev\ and \divis, and ignored the
exponentially suppressed contributions.
Using \dmassl,\heavycouplingsd,\heavymsd,\heavymsplit\ and
\couplings, we can approximate $F$ by
\eqn\fdef{F = {\cal O}(1) { q_2 \over l_1(l_2 + (mL)^{4\nu} l_1)}
\Bigl({mL} \Bigr)^{4\nu} \Bigl(1- \theta(|m|-{q_2 \over l_2})\Bigr)
\, ,}
and upon substituting into \numasn\ we find
\eqn\numasr{m_A^2
\sim  {\cal O}(1){q_2 \over M_5L}{1\over l_1(l_2+(L/l_2)^{4\nu}l_1)} \Bigl({L\over
l_2}\Bigr)^{4\nu} \, .}
Hence this mass is parametrically similar to \squarkmassr.

Now we turn to the gaugino masses induced by the transmission of
SUSY breaking. To lowest order, the gaugino
masses must come from the loop diagram involving a bulk multiplet
with an $R$-symmetry breaking mass term. Because the
gauge fields live only on the brane and not in the bulk, gauge
invariance requires the existence of an adjoint
which couples to the gaugino through terms like
$\sim g Tr(\Phi_b \lambda_b)
\lambda_B$. In order not to spoil gauge unification, the
adjoint must be heavy, $M \sim 1/L$. The adjoint
must also acquire number operator-breaking mass term, in order to
give rise to the gaugino mass. This is clear from
diagram topology, since the ``in"-vertex is generated by the term
$\sim Tr(\Phi_b \lambda_b) \lambda_B$, and the
Majorana mass insertion $m_M \lambda^\dagger_B
\lambda^\dagger_B$ on the $\lambda_B$ line converts
it into $\lambda_B^\dagger$, and so the
``out"-vertex must be
$\sim Tr(\Phi_b \lambda_b) \lambda_B$ to contract this with.
Hence the adjoint must pick up a mass term $\sim \delta m^2_A \Phi_b \Phi_b$
so that the contraction does not vanish.
Such a term arises radiatively,
as given in eq. \numasr.

Next we must
reduce the Majorana mass term to the 4d effective theory.
The left- and right-localized fermions acquire
hierarchically different Majorana masses in four dimensions
but they also
have very different couplings to the
fields on the right wall. Indeed,
for fermions $\lambda_{B~4D} = a^2
\lambda_B$, so the Majorana mass for left-localized
fermions is $a^4 q_2 [\chi^T i \sigma^2 \chi +
h.c.]$=$q_2 [X^T i\sigma^2 X + h.c.]\sim$ $(q_2/l_1) [\psi^T_{4D} i \sigma^2
\psi_{4D} + h.c.]$, \ie\ it is $q_2/l_1$ in the $4D$ theory.
For the right-localized fermions, there is an
additional suppression since $A_L \sim (mL)^{2\nu} A_R$ is the parameter
which sets the scale of the wavefunction on the left wall.
Therefore the right-localized
fermion has tunneling-suppressed $4D$ Majorana mass,
$q_2 f^2(mL)/(l_2 + f^2(mL) l_1)$.
It is small for the light states, but approaches that of the left-localized
bulk fermions for the heavy modes.

Define
$$\tilde q = (q_2/l_1, q_2 f^2(mL) /(l_2 + f^2(mL)l_1))$$
for the left- and right-localized fermions, respectively.
The gaugino mass transmitted to the right IR brane
is, approximately,
\eqn\gaum{
m_{g} \sim {\cal O}(1) \sum_{m} \tilde q \delta m^2_A ~g^2
\int { d^4 p \over (p^2+ M^2)^2 (p^2 + m^2)} \ .}
Here $\delta m^2_A$ is the radiative mass correction of the adjoint,
given in eq. \squarkmassr, and $M \sim L^{-1}$ is the mass of the adjoint.
We sum the series using \sumev. Here
$F = M_5 L \tilde q g^2 \delta m^2_A$. After
doing the angular integrals,
the gaugino mass is approximately
\eqn\gaumsum{\eqalign{
m_{g} & \sim {{\cal O}(1) l \over \alpha M_5L}
 \int^\infty_0
{dp ~p^2 \over (p^2 + M^2)^2} \Bigl({1\over 2} [F(ipL) + F(-ipL)]  \cr
&+
{F(ipL) \over e^{2\pi lp/\alpha - 2i\pi \beta (i pL)}-1} +
{ F(-ipL) \over e^{2\pi lp/\alpha + 2i\pi \beta (- i pL)}-1}
\Bigr)\ .}}
The integral is finite, because it is cut off in the UV
where $F \rightarrow {\rm const.}$ and the integrand is $\sim 1/p^2$.
We can ignore the terms with exponential suppressions in the UV
because the main
contribution to them comes from the momenta $p \sim \alpha/2\pi l$, while
the main contribution to the unsuppressed integrals comes from $p \sim M$,
such that their ratio is $\sim 1/(Ml)^3 <<1$. Thus using the saddle point
method to estimate the integral, we
obtain
\eqn\gaumap{
m_{g} \sim {{\cal O}(1) l \over \alpha M M_5L }
\Bigl(
F({iM L})  +
F(-{i M L}) \Bigr) \ .}
Now, for either the left- or the right-localized modes, the states
with high loop momentum $\sim M$ dominate \gaumap. Since $f^2(ML) \sim 1$, the
relevant bulk fermion Majorana mass is {\it always} $\tilde q = q_2/(l_1 + l_2)
\sim {\cal O}(1) q_2/l_1$. Also, in this limit the couplings approach rapidly those
of eq. \heavycouplings. In particular, the left-localized modes couple to the
right IR brane with $g^2 \sim (M_5L)^{-1} l_2/l_1$, while the right-localized modes
couple more strongly, with $g^2 \sim (M_5L)^{-1}$. This shows clearly that this
integral does not give any tunneling suppression factors
to the gaugino mass, which comes entirely from the tunneling
suppression factors in $\delta m^2_A$. So tunneling suppression terms are one-loop effects.

Therefore, for the left-localized modes, $F=q_2 \delta m^2_A l_2/l_1^2$
and $l=l_1$ \susmasfer,\fercoups, so using \numasr\ for $\delta m^2_A$,
\eqn\gaumapl{
m_{g} \sim {\cal O}(1)  \Bigl({ q_2 \over  M_5L }\Bigr)^2
{1\over M l_1} {1\over l_2 + (L/l_2)^{4\nu} l_1} \Bigl(
{ L \over l_2} \Bigr)^{4\nu}
\ .}
For the right-localized modes, $F = q_2 \delta m^2_A/l_1$
and $l=l_2$ \susmasferr,\fercoups,
and so their contribution to the gaugino mass
is essentially the
same as \gaumapl. Both contributions
are dominated by the states with
loop momenta of the order of the adjoint mass, $p \sim M$.
As before, we have two regimes, depending on $({L /  l_2}
)^{4\nu}l_1/l_2$.

The formula for the gaugino mass generated by
loops of the bulk fields is:
\eqn\gmr{
m_{g} \sim  {\cal O}(1)  \Bigl({ q_2 \over  M_5L }\Bigr)^2
{1\over M l_1}
\times \cases{~
{1\over l_2}
\Bigl({L \over l_2} \Bigr)^{4\nu} ~
& ~~~~~ $l_1 < l_2 \Bigl({  l_2\over L }\Bigr)^{4\nu}$ \cr
~{1 \over l_1 } ~ & ~~~~~
$l_1 > l_2 \Bigl({  l_2\over L }\Bigr)^{4\nu}$  \cr }}
The tunneling suppression is manifest in the form of $(L/l)^{4\nu}$
dependence, and begins to disappear in the limit where
the localization of bulk states is weak.

For comparison with other mediation mechanisms,
we also consider
the corrections to the gaugino mass from gravity and Weyl
anomaly mediation. The gravity mediation would give gaugino mass of
order of teh gravition mass $m_{3/2}$ \suGUT. The anomaly-mediated
supersymmetry breaking leads to a universal form of the gaugino mass \rszero,
$m_q^A = {\beta_g(g^2) \over g^2} m_{3/2}$, where
$g$ is the gauge coupling, $\beta_g$ the gauge coupling
$\beta$-function. The gravitino
mass is generated by the exchange of split bulk multiplets in the
loop. By tree-level Ward identities, the mass formula is,
approximately, \eqn\grms{ m_{3/2} \sim {\cal O}(1) \sum_{m}
{\tilde q \delta m^2_B \over M^2_4} \int d^4p {p^2 \over (p^2 +
m^2)^3} \ ,} where $\delta m^2_B$ is the bulk boson $R$-symmetry
breaking mass. We can again sum over the bulk mode masses using
the technique in \S4.3. Since the $R$-symmetry breaking terms
$\tilde q \delta m^2_B$ depend on the tunneling suppression, the
left-localized mode contributions will dominate. The
left-localized fermion Majorana mass is $\tilde q \sim q_2/l_1$,
while the bulk boson $R$-breaking mass squared is $\delta m^2_B =
F/l_1^2$, where $F$ is ${\cal O}(1)$ for light states and and
decreases with mass as $F \sim q_2/(l_1 m)$ for heavy states. For
simplicity we again approximate it by the step function
$F=1-\theta(|m|-q_2/l_1)$. Using this, noting that
$p^4/(p^2+m^2)^3 = (1/2) \partial^2_s(sp^2 + m^2)^{-1}$ and
recalling \smsonec, where we can now ignore the subleading terms
with exponential momentum suppressions, we find \eqn\sumres{
m_{3/2} \sim {\cal O}(1) {q_2 \over  M_4^2l_1^2} \partial_s^2
s^{-1/2} \int^\infty_0 dp \Bigl(1-\theta(|p|-q_2/l_1)\Bigr)}
Therefore the gravitino mass is, using $l=l_1$ for the
left-localized modes,
\eqn\gravmass{ m_{3/2} \sim {\cal O}(1) {q_2
\over M_4^2 l_1^3}\, .}
 This mode is lighter than
other gravitino KK modes, whose masses are $\sim l_1^{-1},
l_2^{-1}$, because it is protected by SUSY.
So tunneling mediation will be a
dominant mechanism for transmitting SUSY-breaking mass splittings
to the visible sector as long as $m_g > m_{3/2}$. We address the
conditions when this is satisfied below.

\newsec{Phenomenology of Tunneling Mediation}

The SSM cutoff is set by the conformal distance of the SSM to the
UV brane, $l_2^{-1}~\sim~10^{15}~GeV$.  This choice ensures that
there is no fundamental obstacle to implement the supersymmetric
unification of gauge couplings. Squark masses are generated by
radiative corrections including gaugino loops. They start out
close to zero in the UV, and rise via the RG flow in the IR. As a
result, $m_{sq} \sim TeV$ at the electroweak energy and is
comparable to the gaugino mass, as in no-scale models \noscale.

Tunnelling suppression  produces large mass hierarchies without
much effort. For example, take $M_5 \sim 10^{16} GeV$, $L \sim
5/M_5$, $l_2 \sim 5L$, $M\sim 1/L$ and $q_2 \le {\cal O}(1)$. The
tunnelling suppression coefficient $\nu$ and the SUSY breaking
scale $l_1^{-1}$ must be chosen so that $m_g \sim TeV$. For
$\nu=1$, i.e. with little tunnelling suppression, the required
SUSY breaking scale is low, $l_1^{-1} \sim 10^{10} GeV$. This
scale implies a micron-range gravitino mass $m_{3/2} \sim eV$. If
$\nu=3$, the SUSY breaking scale should be $l^{-1}_1 \sim 3\times
10^{13} GeV$, closer to the unification scale. The induced
gravitino mass is $m_{3/2} \sim 270 GeV$.

Does tunneling mediation solve the flavor problem? Just as in
gravity mediation, the answer is, in general, negative. The reason
is that although the squark masses vanish at precisely the GUT
scale, they are non-zero an order of magnitude below $M_{GUT}$;
the squark masses are ``hard" --they do not vanish sufficiently
rapidly in the UV. So they can be distorted by the nearby flavor
physics,  taking place at $M_{GUT}$ or $M_{Pl}$, which is also
responsible for the large  intergenerational differences of quark
and lepton masses. These distortions can cause a misalignment of
squark and quark masses leading to unacceptable flavor violation.
The only way to avoid this is for the UV theory to be special,
e.g. the squark masses could line up with the quark masses and
therefore not create new flavor violating angles.

There are also model-independent gravity \refs{\hs,\suGUT} and anomaly \rszero\
mediated contributions to the sparticle masses that are bounded by $m_{3/2}$.
They are subdominant to the tunneling mediated contributions
as long as
\eqn\ineq{l_1 > l_2 \bigl({l_2 \over L}
\bigr)^{2\nu + 1/2} {1 \over M_4l_2}\, .}
It is not difficult to satisfy this constraint because of the
4d Planck mass $M_4$ in the denominator on the RHS.

The hierarchies we produce do not originate from the AdS scaling
as in \RSI. In our case the cutoff on the SSM brane is $M_{GUT}$.
Furthermore, our effect would persist with slight modifications
given any warp factor which raises a barrier between different
throats.



\newsec{Tunneling, 4D Effective Field Theory, and Compactified AdS/CFT}

In this section we will discuss the four-dimensional description
of our results, in the case that the
throats are the near-horizon geometries of
D3-branes.   In this background,
normalizable closed string excitations
have a dual description as the massless gauge theory excitations
on the D3-brane worldvolume \juan.  More precisely, if one does
not scale $\ap$ strictly to zero, there are
scattering states in the asymptotically flat region
which are purely ingoing at the AdS horizon;
and modes which are localized in the AdS throat,
outgoing at the past horizon, and ingoing at the future horizon
\refs{\bgl,\scatt}. In
the strict near horizon limit with $\ap\to 0$,
the latter become normalizable modes in AdS, which are
dual to states of the D-brane CFT created
by modes of local gauge-invariant operators
${\cal O}_I(x)$ \refs{\albion,\bklt,\bdhm}.
The former become non-normalizable modes
which are dual to couplings $\lambda_I$
multiplying these operators \refs{\AdSCFT,\albion}\ in the dual
CFT description:
$$
    \int d^4 x \lambda_I {\cal O}_I (x)\ .
$$
We will refer to these modes as ``asymptotic closed strings'' in
reference to the localization of their wavefunctions in the
asymptotically flat region of the full D3-brane geometry.

We will focus on the five-dimensional
graviton $h_{\mu\nu}$.  When $h$ is polarized along
the boundary, the normalizable and non-normalizable
modes of $h$ are, respectively,
dual to the modes of and couplings multiplying
the gauge theory stress tensor $T_{\mu\nu}$.

When the AdS throat opens into some compact manifold,
the low-energy states supported in this manifold
become modes of four-dimensional (super)gravity plus additional
fields arising from the compactification.  These
modes can be modeled as excitations on the UV (``Planck'') brane,
which couple to modes supported in the bulk of AdS \HV.
We might expect to maintain the interpretation of these
latter modes as excitations of a dual CFT with a cutoff,
now coupled to four-dimensional (super)gravity \RSAdS.
In the setup of \S2, the low-energy action would then
include arbitrary products of gauge theory operators
\eqn\newops{
    \delta S = \int d^4 x \lambda_{ij} \CO^{(1)}_i
        \CO^{(2)}_j + \ldots\ ,
}
where the superscripts $(1,2)$ label the left and right
AdS regions of Fig. 2.

The tunneling calculation in \S2\ tests this
proposal.  An operator $\CO$ of dimension $\Delta$
will appear in the action with powers of the UV cutoff
$\luv$:
\eqn\eff{
    \int d^4 x \luv^{4-\Delta} \lambda \CO_\Delta
}
where $\lambda$ is dimensionless.
If $\CO_\Delta$ takes the form \newops,
it will mediate transitions between modes
of these operators with a probability:
\eqn\gentrans{
    P \sim \lambda \left(\frac{m}{\luv}\right)^{2\Delta - 8}
}
and a transition rate:
\eqn\genrate{
    \Gamma \sim m \lambda
    \left(\frac{m}{\luv}\right)^{2\Delta - 8}
}
where $m$ is the mass of the initial state.
The results of \S2\ for the tunneling of graviton
modes give:
\eqn\Pans{
 P\sim m^4L^4\sim e^{-4R_1/L} }
and
\eqn\rate{ \Gamma\sim {1\over
L}e^{-5R_1/L} . }\

A glueball near the left IR brane has mass:
\eqn\gluemass{
    m \sim \frac{1}{L} e^{-R_1/L}\ .
}
We conclude that the transition between glueballs
is mediated by a dimension-6
operator and that the natural UV cutoff
is $\luv = 1/L=M_{UV}$.\foot{There is some freedom in
this identification of the cutoff.  In the denominator
of \metricL\ we can replace $L$ with $\zeta L$.
The tunnelling formula is the same, but with
$\Lambda_{UV} = 1/\zeta L$.  $\zeta$ will
be fixed by the details of the microscopic model.}
We note first that this is not the four- (or five- or ten-) dimensional
Planck scale as would appear in the minimal coupling of
gravity to a CFT.  Instead,
the dimensionful parameters are set by scales in the
background geometry.\foot{In a perturbative string limit,
$M_s < M_{(pl)4,5,10}$.  If the throats were close enough,
the strings stretched between the D-branes could play an
appreciable role in the dynamics below the
Planck scale.  In the case at hand
these ``W bosons'' are much heavier than $M_s$.}

A further surprise is that
the operator mediating the transition in \S2\
is not constructed from the CFTs alone.
The $2\times({\rm CFT})$
operator mediating a transition between
the gravitons $h_{\mu\nu}$ in the two AdS regions
would be the dimension {\it eight}\ operator:
$$ :T_{\mu\nu}^{(1)}T_{\mu\nu}^{(2)}: $$
However, the outgoing wave
in the second throat has a wavefunction proportional to
\eqn\onecoeff{\psi
~\sim~ J_{2} ~+~i~N_{2}\ .}
In the AdS/CFT dictionary this is a Fourier
mode of the ``bulk-boundary propagator'' \refs{\AdSCFT,\bgl}\
and corresponds to a change in the coupling, i.e. an
asymptotic closed string mode,
$$\delta h_{\mu\nu} T_{\mu\nu} \ .$$
In field theory language, the calculation of \S2.1\ is computing
the overlap
\eqn\ftcomp{\langle \psi_1 \vert T^{(1)}_{\mu\nu}~ H^{(2)~\mu\nu}
~\vert \psi_2
\rangle}
$\vert \psi_1 \rangle$ is the state on the left side of the wall,
and $\vert \psi_2\rangle$ is the state with outgoing
wavefunction $J + iN$ on the right.  If this
overlap is to be non-vanishing,
$H^{(2)}$ must be built from asymptotic
closed-string degrees of freedom.

There are many candidates for $H$.
Obvious terms constructed from bulk gravitons are:
$R^{(5)}_{\mu\nu}(z=0,x)$ and $g^{(5)}_{\mu\nu}R^{(5)}(z=0,x)$,
where these fields are now included in the 4d effective action.
In addition, the ``UV wall'' defines
a covariant normal vector $n_\mu$,
whose covariant derivatives can be combined into an infinite
number of dimension-two covariant two-index tensors.  Two sets
of terms are $(n\cdot\nabla)^\alpha K_{\mu\nu}(x)$
and $\nabla_\mu\nabla_\nu(n\cdot\nabla)^\alpha K$,
where $K_{\mu\nu}$ is the extrinsic curvature of the UV wall.

Another sign that we must couple the CFT and asymptotic
closed string operators is that in
the standard RS picture, the mass eigenstates
for $h$ take the form:
\eqn\smallcoeff{
    \psi
    ~\sim~ J_{2} ~+~b~(mL)^2~N_{2}\ ,
}
where $b$ is of order 1.
When $L\to 0$, $J_2$ and $N_2$ become normalizable and
non-normalizable, respectively.  This implies that the
stress tensor is modified:
\eqn\TT{
    \tilde{T}_{\mu\nu} = T^{{\rm CFT}}_{\mu\nu}
        + b \left(L^2 \p^2\right)\frac{1}{L^2}
        I_{\mu\nu}
}
where $I$ is a linear combination of dimension-two
tensors constructed from the 10d graviton polarized
along the D3-brane.  The additional
derivatives reflect the coefficient of $N$
in \smallcoeff.  Such a mixing will generically occur for
all other operator modes as well.  So the Hamiltonian
mixes the CFT and asymptotic closed-string degrees of freedom;
the effective theory includes operators from both sectors.

The small, derivative-suppressed coefficient
in \smallcoeff,\TT\ also suppresses
the dimension-four operator one might expect from
the purely asymptotic closed-string contribution to
$:\tilde{T}^{(1)}\tilde{T}^{(2)}:$.
This derivative suppression also
means that at sufficiently low energies the
modes localized in one AdS throat
are well described by the original CFT
stress tensor.  However, at best this decoupling works
when studying a single throat.  When
coupling the two, we must include modes
from both throats, and asymptotic closed string modes,
even at low energy.
We must take care with the scale-radius correspondence when
the AdS throat is coupled to additional degrees of
freedom.  Low energy modes are supported in the compactification
manifold or down the right throat as well as far down the
left throat.
Furthermore, if one chooses boundary conditions such that
wavefunctions have larger mixings between $J$ and $N$,
even the dynamics far down one AdS throat will cease to
be well described by CFT dynamics.

The perturbations forced on us are distinct from another
effect we might expect.  A conformal field theory coupled
to gravity will generically have all relevant (and irrelevant)
CFT operators appearing in the Lagrangian, with coefficients
given by suitable powers of the UV scale $M_4$.
This is $\it not$ the effect we are discussing here.  As in all
discussions of Randall-Sundrum scenarios, we have assumed
the absence of any such relevant operators (tachyons satisfying
the Breitenlohner-Freedman bound \refs{\brfreed,\AdSCFT});
these would yield an instability of the gravity solution after introducing
the UV brane.

\vskip1cm
\centerline{\bf{Acknowledgements}}

It is a pleasure to thank N. Arkani-Hamed, S. Giddings, J.
Maldacena, A. Peet, M. Peskin, J. Polchinski, A. Pomarol,
M. Schmaltz, S.
Shenker, P. Steinhardt, A. Strominger, L. Susskind, S. Thomas, H.
Tye and H. Verlinde for interesting discussions about related
subjects. This work was supported in part by the DOE under
contract DE-AC03-76SF00515. During the completion of this project,
the authors enjoyed the hospitality of the Institute for
Theoretical Physics at Santa Barbara, and were supported by the
National Science Foundation under grant number PHY-99-07949.  A.L.
would like to thank the Center for Geometry and Theoretical
Physics at Duke University for its hospitality while this work was
completed; he received additional support there from NSF grant
DMS-0074072. The work of S.K. was supported in part by a
Packard Fellowship for Science and Engineering
and an Alfred P. Sloan Foundation Fellowship, and the work of E.S.
was supported in part by a DOE OJI grant and an Alfred P. Sloan
Foundation Fellowship.

\listrefs

\end